\documentclass[11pt,a4paper]{article}
\usepackage{jheppub}
\usepackage{tikz}

\usepackage{graphicx,verbatim}
\usepackage{amsmath}
\usepackage{amsfonts}
\usepackage{amssymb}
\usepackage{psfrag}
\usepackage{accents}
\usepackage{mathrsfs}
\usepackage[colorinlistoftodos]{todonotes}
\usepackage{epstopdf}

\def\be{\begin{equation}}
\def\ee{\end{equation}}
\def\ba{\begin{eqnarray}}
\def\ea{\end{eqnarray}}

\def\={\hat{=}}

\def\f{\frac}

\def\rmd{\mathrm{d}}
\def\Sbb{\mathbb{S}}
\def\Rbb{\mathbb{R}}

\def\Lie{\mathcal{L}}
\def\S{\mathcal{S}}
\def\T{\mathcal{T}}

\def\H{\mathcal{H}}
\def\B{\mathcal{B}}

\newcommand{\pb}[1]{\hbox{\lower0.8ex\hbox{${}_{\leftarrow}$}}\kern-1.9ex{#1}}

\def\G{\mathfrak{G}}
\def\BMS{\mathfrak{B}}
\def\b{\mathfrak{b}}
\def\s{\mathfrak{s}}
\def\g{\mathfrak{g}}

\def\Spi{\mathfrak{S}}

\def\L{\mathfrak{L}}

\def\qo{\mathring{q}}
\def\go{\mathring{g}}
\def\Co{\mathring{c}}
\def\Ko{\mathring{k}}

\def\Do{\mathring{D}}
\def\vac{\mathring{\mathfrak{D}}}
\def\no{\mathring{n}}
\def\mo{\mathring{m}}
\def\ello{\mathring{\ell}}
\def\sigmao{\mathring{\sigma}}

\def\omegao{\mathring{\omega}}
\def\epsilono{\mathring{\epsilon}}

\def\SO(3){\rm SO(3)}
\def\so(3){\rm so(3)}
\def\SO(4){\rm SO(4)}
\def\so(4){\rm so(4)}
\def\SO(1,4){\rm SO(1,4)}
\def\so(1,4){\rm so(1,4)}
\def\SU(2){\rm SU(2)}

\def\hM{\hat{M}}
\def\hg{\hat{g}}

\def\rmd{\mathrm{d}}

\def\S{\mathcal{S}}

\def\D{\mathcal{D}}

\newcommand*{\scri}{\ensuremath{\mathscr{I}}} 
\newcommand*{\scrip}{\ensuremath{\mathscr{I}^{+}}} 
\newcommand*{\scrim}{\ensuremath{\mathscr{I}^{-}}} 
\newcommand*{\scripm}{\ensuremath{\mathscr{I}^{\pm}}} 
\def\inot{i^\circ}
\def\Poincare{\rm Poincar\'e\,\,}
\def\bsP{P_{\!a}^{\rm BS}}% 4-momentum
\def\admP{P_{\!a}^{\rm ADM}}% 4-momentum
\def\inotP{\mathfrak{p}_{\inot}}% 4-momentum
\def\bmsP{\mathfrak{p}^{\rm bms}_{\inot}} %Poincare subgroup
\def\spiP{\mathfrak{p}^{\rm spi}_{\inot}} % Poincare subgroup
\def\scriJ{\vec{J}_{\!\scrip}}
\def\inotJ{\vec{J}_{\inot}}

\def\Omegao{\mathring{\Omega}}
\def\go{\mathring{g}}
\def\nablao{\mathring{\nabla}}
\def\no{\mathring{n}}
\def\lo{\mathring{\ell}}
\def\Co{\mathring{C}}
\def\Ko{\mathring{K}}
\def\uo{\mathring{u}}
\def\No{\mathring{N}}

\def\So{\mathring{S}}
\def\Ro{\mathring{R}}
\def\alphao{\mathring{\alpha}}
\def\muo{\mathring{\mu}}
\def\fo{\mathring{f}}

\def\ubS{\underbar{S}}
\def\ubSo{\mathring{\underbar{S}}}

\def\bfB{\mathbf{B}}
\def\bfD{\mathbf{D}}
\def\bfE{\mathbf{E}}
\def\bff{\mathbf{f}}

\def\bfh{\mathbf{h}}
\def\bfK{\mathbf{K}}

\def\bfS{\mathbf{S}}
\def\bfT{\mathbf{T}}
\def\bfalpha{\boldsymbol{\alpha}}
\def\bfbeta{\boldsymbol{\beta}}
\def\bfepsilon{\boldsymbol{\epsilon}}
\def\bfphi{\boldsymbol{\phi}}
\def\bfzeta{\boldsymbol{\zeta}}

\begin{document}
\title{Unified Treatment of Null and Spatial Infinity III: Asymptotically Minkowski Space-times}
\author[a]{Abhay Ashtekar,}
\affiliation[a]{Physics Department \& The Institute for Gravitation and the Cosmos, Penn State, University Park, PA 16802, U.S.A.\\
Perimeter Institute for Theoretical Physics, 31 Caroline St N, Waterloo, ON N2L 2Y5, Canada}
\emailAdd{ashtekar.gravity@gmail.com}
\author[b]{Neev Khera}
\affiliation [b]{Physics Department, University of Guelph, Guelph, Ontario, N1G 2W1, Canada}
\emailAdd{nkhera@uoguelph.ca}

\abstract{
\noindent

The Spi framework provides a 4-dimensional approach to investigate the asymptotic properties of gravitational fields as one recedes from isolated systems in any space-like direction, without reference to a Cauchy surface \cite{aarh}. It is well suited to unify descriptions at null and spatial infinity because $\scri$ arises as the null cone of $\inot$. The goal of this work is to complete this task by introducing a natural extension of the asymptotic conditions at null and spatial infinity of \cite{aa-ein}, by `gluing' the two descriptions appropriately. Space-times satisfying these conditions are asymptotically flat in both regimes and thus represent isolated gravitating systems. They will be said to be \emph{Asymptotically Minkowskian} at $\inot$. We show that in these space-times the Spi group $\Spi$ as well as the BMS group $\B$ naturally reduce to a single \Poincare group, denoted by $\inotP$ to highlight the fact that it arises from the gluing procedure at $\inot$. The asymptotic conditions are sufficiently weak to allow for the possibility that the Newman-Penrose component $\Psi^\circ_1$ diverges in the distant past along $\scrip$. This can occur in astrophysical sources that are not asymptotically stationary in the past, e.g. in scattering situations. Nonetheless, as we show in the companion paper \cite{ak-J}, the energy momentum and angular momentum defined at $\inot$ equals the sum of that defined at a cross-section $C$ of $\scrip$ and corresponding flux across $\scrip$ to the past of $C$, when the quantities refer to the preferred \Poincare subgroup $\inotP$.}

%\pacs{04.20.Cu, 04.60.+n,04.30.+n}
%\pacs{04.70.Bw, 04.25.dg, 04.20.Cv}
%\pacs{98.80.Qc, 04.60.Pp, 04.60.Kz}
\maketitle

\section{Introduction}
\label{s1}

Isolated gravitating systems in general relativity are represented by asymptotically flat space-times. In these space-times, curvature falls off as one recedes from sources either along null or space-like directions. Receding in null directions is especially convenient in the analysis of the radiative aspects of the gravitational (and other zero rest mass) fields. This analysis was carried out by Bondi, Sachs et al. \cite{bondi,sachs} and was recast in a more geometric setting using a conformal completion by Penrose \cite{rp}. The framework brought out a surprising fact that, in presence of gravitational waves, the asymptotic symmetry group at null infinity is not the \Poincare group, as one might expect at first, but an infinite dimensional generalization thereof, the Bondi-Metzner-Sachs (BMS) group $\BMS$. The BMS group does admit a canonical 4-dimensional normal subgroup $\T$ of translations, whence there is a well-defined notion of Bondi-Sachs 4-momentum ${\bsP} [C]$ at a retarded instant of time represented by a cross-section $C$ of $\scrip$. However, whereas the \Poincare group admits only a 4-parameter family of Lorentz subgroups, the BMS group admits an infinite parameter family, labelled by a general (conformally weighted) function on a 2-sphere. As a result, the notion of angular momentum at null infinity acquires a supertranslation ambiguity. However, it has been known for quite some time \cite{etnrp,aa-rad} that by imposing an additional---but rather tame---boundary condition on the ``magnetic part" of Weyl curvature as one approaches $\inot$ \emph{along} $\scrip$, one can extract a canonical \Poincare subgroup $\bmsP$ of the BMS group $\BMS$ (which, in Minkowski space, coincides with its isometry group).%
\footnote{Electric and magnetic parts of the Weyl tensor normally refer to a decomposition using time-like normals to space-like hypersurfaces. At null infinity, one uses instead the \emph{null} normal to $\scrip$ and at spatial infinity the \emph{space-like} normal to the unit (time-like) hyperboloid at $\inot$. Hence, the use of quotes in the phrase ``magnetic part".}
Thus, by supplementing asymptotic conditions as one approaches $\inot$ in \emph{null} directions, one can remove the supertranslation ambiguity and arrive at a definition angular momentum at $\scrip$ as in special relativity. Since this reduction refers to $\inot$, it is the appropriate framework for comparison with the total angular momentum defined by approaching $\inot$ in space-like directions \cite{aarh,aa-ein}.

The asymptotic structure of the gravitational field as one recedes from sources in space-like directions was first analyzed by Arnowitt, Deser and Misner (ADM) in the 3-dimensional setting, by focusing on the behavior of initial data on (partial) Cauchy surfaces $\Sigma$ (for a summary, see, e.g., \cite{adm}). Later, Geroch \cite{rg-jmp} analyzed this structure using a conformal completion of $\Sigma$. This reformulation brought out the fact that there is again a supertranslation freedom that had been largely ignored in the earlier treatments. However, subsequently it was shown that this freedom can also be removed by imposing an additional, mild fall-off condition on the spatial curvature of the 3-metric of the 3+1 setting \cite{am3+1}. This condition is the restriction to the 3+1 framework of a condition, again on the ``magnetic part" of the Weyl tensor but now along spatial directions \cite{aarh,aa-ein}. For this class of initial data, the asymptotic symmetry group on Cauchy slice reduces to the Euclidean group and the notion of angular momentum at spatial infinity is free of the supertranslation ambiguity. Its expression agrees with the one given in the ADM framework, provided of-course, one uses the rotational symmetries that belong to the preferred Euclidean subgroup of the asymptotic symmetry group. For descriptions of spatial infinity \emph{by itself} from a 4-dimensional perspective, see, e.g., \cite{rbbs,aajr} for approaches that use asymptotic field equations and \cite{henneaux2,henneaux3} for a more recent framework based on 3+1 Hamiltonian methods that discusses the BMS group in the context of spatial infinity.

To relate physics at null infinity to that at spatial infinity, one needs a 4-dimensional generalization of the ADM framework that also includes $\scripm$. Basic ideas necessary for this unification were introduced in \cite{aarh} and the final conceptual framework appeared in \cite{aa-ein}. This paper is a continuation of those two investigations. We will follow the treatment given in \cite{aa-ein} because it brings to forefront the causal structure, the conformal metric, and its torsion-free connection, that are more directly useful in extracting physics from asymptotic geometry.%
\footnote{For a PDE/geometric analysis perspective on this unification, see in particular \cite{hf,kroon1,kroon2}, and for more recent works that use conceptually different frameworks, see \cite{kpis,Prabhu:2019fsp,cgw}.} 
Now $\scripm$ arise as the future and the past light cone of $\inot$ of the conformally completed space-time. However, since the immediate goal of that analysis was to construct a 4-d generalization of the ADM framework, it focused on the approach to $\inot$ \emph{only in space-like directions}. In this setting the asymptotic symmetry group---called the `Spi group' \cite{aarh,aa-ein} and denoted here by $\Spi$--- is again an infinite dimensional generalization of the \Poincare group as in the BMS case. However, supertranslations are now labelled by functions on the \emph{3-dimensional} unit hyperboloid $\H$ in the tangent space at $\inot$. One can again impose an additional asymptotic condition on the ``magnetic part" of the Weyl tensor---now in the approach to $\inot$ in arbitrary \emph{space-like} directions---and reduce $\Spi$ to a canonical \Poincare subgroup  $\spiP$. Then the total relativistic angular momentum of the system can again be defined without supertranslation ambiguities. Thus, the overall situation at null infinity $\scrip$ and at spatial infinity $\inot$ is similar. 

Therefore, a question naturally arises: Is there a \emph{natural} geometric `gluing' of the boundary conditions used along space-like and null approach to $\inot$ that lead to the \Poincare \\reductions of both $\BMS$ and $\G$ in one stroke? It is important of course that the gluing conditions should not rule out physically interesting situations. In this paper we will answer this question in the affirmative. 

As noted above, the emphasis in \cite{aarh,aa-ein} was on the asymptotic behavior of fields as one approaches $\inot$ along \emph{space-like directions}. We will extend that analysis by including natural continuity requirements on the approach to $\inot$ along space-like \emph{and} null directions. Space-times satisfying these conditions will be referred to as being \emph{Asymptotically Minkowsian (AM)}. We will find that, in these space-times, the condition (and the subsequent procedure) used at $\scrip$ is in fact just a continuous extension of the condition used in space-like directions. Thus, as one would intuitively expect, asymptotic symmetries in \emph{AM} space-times constitute a (single) \Poincare group $\inotP$. Restriction of these symmetries to $\scrip$ yields the subgroup $\bmsP$ of $\BMS$, and restriction to spatial infinity yield the \Poincare subgroup $\spiP$ of $\Spi$. Given a generator $\xi^a$ of $\inotP$ the standard framework at null infinity provides a charge $Q_\xi [C]$ for each cross-section $C$ of $\scrip$ and the one at spatial infinity provides a charge $Q_\xi [\inot]$. In the companion paper we will show that the physically expected relation holds: $Q_\xi [\inot]$ is sum of $Q_\xi[C]$ and the flux $F_\xi [\inot,C]$ across the portion of $\scrip$ to the past of $C$. This will follow from the definition of \emph{AM} space-time alone. 
Our gluing procedure has some similarities with \cite{Prabhu:2019fsp, kpis} but, as our discussion of sections \ref{s3} and \ref{s4} shows, there are also key conceptual and technical differences. These are discussed further in section 4 of \cite{ak-J}.

The material is organized as follows. To make the discussion reasonably self-contained, in section \ref{s2} we recall the structures at null and spatial infinity that are essential to our analysis in this paper and its follow-up \cite{ak-J}. In particular, we explain why---although the \Poincare reduction $\bmsP$ the BMS group $\B$ has been known for a long time \cite{etnrp}---it is not used in the literature that focuses on null infinity alone. In section \ref{s3} we introduce the notion of \emph{AM} space-times by introducing conditions that `glue' the behavior of various fields in the approach to $\inot$ along space-like directions and along $\scrip$. In section \ref{s4} we show that the asymptotic symmetry group for these space-times is just a single \Poincare group $\inotP$ whose restriction to the two regimes coincides with the \Poincare subgroups $\bmsP$ of $\BMS$ and $\spiP$ of $\Spi$ that were isolated separately. In section \ref{s5} we summarize the results and put them in a broader context. In particular, we explain why, although the \Poincare group $\inotP$ is the appropriate symmetry group for isolated systems in classical general relativity, this fact does not reduce the significance of the BMS group which has important applications in both classical and quantum gravity.

Our conventions are the following. The underlying space-time is 4-dimensional. The physical metric is denoted by $\hat{g}_{ab}$ and the conformal metric by $g_{ab}$. We use  -,+,+,+ signature. The torsion-free derivative operator of $g_{ab}$ is denoted by $\nabla$ and curvature tensors are defined via:  $2\nabla_{[a}\nabla_{b]} k_c = R_{abc}{}^d k_d$, $R_{ac} = R_{abc}{}^b$, and $R=g^{ab}R_{ab}$.  In case of an ambiguity, will use the equality \, $\=$\, to emphasize that equality that holds only at $\scrip$ or $\inot$. 

\section{Structures at null and spatial infinity}
\label{s2}

In this section we recall the asymptotic structures and fields using the framework that will be used in our analysis. This discussion will also enable us to fix the notation. For further details, see, e.g., review articles \cite{rg,frau,aa-yau,adm,aa-padova,henneaux3}.

\subsection{Null infinity $\scrip$}
\label{s2.1}

Let us begin with the notion of asymptotic flatness at null infinity \cite{aabs} that is most suitable for the purposes of this paper. For concreteness, we will focus on $\scrip$.\\
\noindent  \textbf{Definition 1:} A physical space-time $(\hat{M},\hat{g}_{ab})$ will be said to be \textit{asymptotically flat  at null infinity} if there exists a manifold $M$ with boundary $\scrip$ equipped with a $C^4$ metric $g_{ab}$ and a diffeomorphism from $\hat{M}$ onto $M\, \setminus\, \scrip$ (with which we identify $\hat{M}$ and $M\, \setminus\, \scrip$) such that:

i)\, there exists a smooth function $\Omega$ on $M$ with $g_{ab} = \Omega^2 \,\hat{g}_{ab}$ on $\hat{M}$;\,\, $\Omega=0$ on $\scrip$;\\
\indent\indent and $n_a := \nabla_a \Omega$ is nowhere vanishing on $\scrip$;

ii) $\scrip$ is topologically $\Sbb^2\times \Rbb$;\,\,\, and,

iii)\,$\hat{g}_{ab}$ satisfies vacuum Einstein's equations $\hat{R}_{ab} - \frac{1}{2} \hat{R} \hat{g}_{ab}  = 0$ in a neighborhood of $\scrip$.\vskip0.1cm

The first condition ensures that the boundary $\scrip$ is at infinity with respect to the physical metric $\hat{g}_{ab}$ and the conformal factor falls-off at an appropriate rate ($\sim ``1/{r}''$). Since $\nabla_{a}\Omega \not=0$ on $\scrip$, $\Omega$ can be used as a coordinate on $M$; we can perform Taylor expansions in $\Omega$ to capture the degree of fall-off of physical fields (reflecting the ``$1/{r}$ expansion'' in the physical space-time). The topological restriction on $\scrip$ captures the idea that one can move away from the isolated system along null rays in any angular direction. The last condition can be weakened to allow matter fields in a neighborhood of $\scrip$: the standard results go through if  $\Omega^{-2} \hat{T}_{ab}$ has a $C^2$ limit to $\scrip$. However, we will assume condition iii) for simplicity. Finally, while we assume $C^4$ differentiability on $g_{ab}$ in the main body of the paper to simplify the discussion, as discussed in {section \ref{s5},} main results go through under weaker conditions.

Conditions i) and iii) immediately imply that $\scrip$ is necessarily a null, 3-dimensional manifold; $n^a = \nabla^{a}\Omega$ is null at $\scrip$ (but not on $\hat{M}$). Thus, $\scrip$ is ruled by the integral curves of $n^a$, which we will refer to as the \emph{generators} of $\scrip$. Condition ii) implies that the space $\bar{S}$ of generators is topologically $\Sbb^2$. We will denote the pull-back of the conformally rescaled metric to $\scrip$ by $q_{ab}$; it has signature 0,+,+ and is the lift to $\scrip$ of a positive definite metric $\bar{q}_{ab}$ on $\bar{S}$.

Next, note that there is freedom to perform conformal rescalings: If $\Omega$ is a permissible conformal factor for a physical space-time $(\hat{M}, \hat{g}_{ab})$, so is $\Omega^\prime = \omega \Omega$ where $\omega$ is smooth on $M$ and nowhere vanishing on $\scrip$. Using this freedom, one can always choose a conformal completion such that $n^a$ is \emph{divergence-free} on $\scrip$. This turns out to be an extremely convenient choice in the discussion of null infinity by itself. We will denote the conformal factors that make $\scrip$ divergence-free by $\Omegao$,\, the corresponding conformally rescaled metric by $\go_{ab}$ (so that $\go_{ab} = \Omegao^2 \hat{g}_{ab}$), and the 1-form $\nablao_a \Omegao$ by $\no_a$.\, (The remaining restricted conformal freedom is given by ${\Omegao}^\prime = \muo\,\Omegao$ where $\Lie_{\mathring{n}} \muo\, =\, 0$ on $\scrip$.)\, Field equations imply that $\nabla_a n_b$ is proportional to $g_{ab}$ at $\scrip$ in any conformal frame, whence in our divergence-free conformal frames we have
\be \label{divfree} \nablao_a \no_b = \nablao_a \nablao_b \Omegao \,\,\= \,\,0\ee 
where, as noted before,\, $\=$\, denotes equality at points of $\scrip$. Eq. (\ref{divfree}) implies that the torsion-free derivative operator $\nablao_a$ compatible with $\go_{ab}$ induces a torsion-free derivative operator $\Do$ defined intrinsically on the 3-manifold $\scrip$ satisfying (\ref{divfree}) 
\be \label{Do} \Do_a \qo_{bc}\, \=\, 0, \quad {\rm and}\quad \Do_a \no^b\,\, \=\,\, 0\, .
\ee
 Although $\qo_{ab}$ constrains $\Do$, it does not determine $\Do$ because it is degenerate. In particular, under $\Omegao \to \muo\Omegao$ on $M$, $\Do$ changes even when $\muo = 1 + \Omegao f$ because $\nabla$ is sensitive to the first order change in $\Omegao$ away from $\scrip$. Therefore, one is led to consider an equivalence class $\{\Do\}$ of intrinsic connections $\Do$ on $\scrip$ where two are regarded as being equivalent if they arise from this trivial change in the conformal factor with $\mu\, \= \,1$ on $\scri$. This implies that $\Do \approx \Do^{\prime}$ if and only if, for any 1-form $K_a$ on $\scrip$,\, $(\Do^\prime_a - \Do_a)k_b = f \,\qo_{ab} \,\no^c k_c$ for some function $f$. Bondi news $\No_{ab}$ is encoded in the curvature of $\{\Do\}$ \cite{aa-rad,aa-bib}.

Finally, \emph{Definition 1} implies that the Weyl tensor $\Co_{abc}{}^d$ of $g_{ab}$ vanishes on $\scrip$. Therefore, the tensor field
\be \label{Ko} \Ko_{abc}{}^d := \Omega^{-1}\, \Co_{abc}{}^d  \ee
has a smooth limit to $\scrip$; it represents the \emph{leading order Weyl curvature} at $\scrip$. The equivalence class of intrinsic derivative operators $\{\Do\}$ on $\scrip$ suffices \cite{aa-rad,aa-bib} to determine the ``magnetic part" 
\be {}^\star\!{\Ko}_{ac}\, := \, {}^\star\Ko_{abcd}\,\, \no^b\,\no^d \, \equiv\,  \textstyle{\f{1}{2}} \mathring\epsilon_{ab}{}^{mn}\,\Ko_{mncd}\, \no^b\,\no^d \ee
of $\Ko_{abcd}$, where the ``magnetic part" is defined using the null normal to $\scrip$.% 
\footnote{In contrast to references \cite{rg,aa-rad,aa-bib,aams}, we will relate limits of fields in the approach to $\inot$ along null {\emph and space-like} directions. Therefore, in this paper we regard tensor fields defined at points of $\scrip$ as living on the 4-manifold $M$, rather than being defined intrinsically on the 3-manifold $\scrip$. Thus, indices are raised and lowered using the 4-metric $\go_{ab}$ rather than the intrinsic metric on $\scrip$, e.g., ${}^\star{\Ko}_{ac} = \go_{ab}\go_{cd}\, {}^\star{\Ko}^{bd}$.}
The Schouten tensor $\So_{ab} := \Ro_{ab} - \textstyle{\f{1}{6}}\, \Ro\, \go_{ab}$ serves as a potential of 
${}^\star\!{\Ko}_{ac}$: 
\be \label{schouten} {}^\star\!{\Ko}_{ab} = 2\, \epsilono_{macd}\, \no^m\,\, \Do^c \So_{b}{}^d. \ee 
This property will be useful in section \ref{s3}. The Bondi News tensor $\No_{ab}$ is the ``conformally invariant part" of the trace-free portion of $\So_{ab}$ \cite{rg}. One can show that if ${}^\star\!{\Ko}_{ac}\, \=\,0$ then $\No_{ab}\, \=0\,$ \cite{abk1}. All connections in the equivalence class $\{\Do\}$ have the same ${}^\star\!{\Ko}_{ac}$ and $\No_{ab}$. Finally, in the Newman-Penrose notation, the 5 components of\, ${}^\star\!{\Ko}_{ac}$ correspond to $\Psi_4^\circ,\, \Psi_3^\circ$ and $\rm{Im}\, \Psi_2^\circ$. They carry only the ``radiative information" in the asymptotic curvature but not the ``Coulombic information" which is carried, e.g., in $\rm{Re}\, \Psi_2^\circ$ and $\Psi_1^\circ$. Thus, ${}^\star\!{\Ko}_{ac}\, \=\,0$  implies absence of gravitational radiation; in particular the equality holds in stationary space-times. 
\goodbreak

\emph{Remarks:}

1. In the discussion of black holes, it is important to ensure that $\scrip$ is complete in the sense that the null normal $\no^a$ is a complete vector field \cite{rggh}. (If it is complete in one divergence-free conformal frame, then it is complete in any other divergence-free frame.) In the discussion of the relation between structures at $\scrip$ with those at $\inot$ it suffices to require that $\no^a$ be \emph{past complete}, i.e., that the affine parameter $\uo$ of $\no^a$  (defined by $\no^a \Do_a \uo =1$) extends to $-\infty$ in the distant past along $\scrip$. There are examples of boost-symmetric space-times \cite{aatd,jbbs} in which $\scrip$ is past complete but not future complete. (Note also that although to have a well-defined action of the BMS group, it is necessary that $\scrip$ be both past and future complete, in the discussion of the BMS Lie algebra it is not.)  \vskip0.1cm

2. Given a null tetrad \`a la Newman and Penrose \cite{rp,rpwr}, the five complex functions $\Psi_4^\circ, \ldots \Psi_0^\circ$ capture the 10 components of $\Ko_{abc}{}^d$ at $\scrip$. The `peeling' property of these 5 NP scalars is a straightforward consequence of the continuity of $\Ko_{abc}{}^d$ at $\scrip$ and the relation between the null tetrads defined by $\hat{g}_{ab}$ and $\go_{ab}$. If $\go_{ab}$ is $C^4$ at $\scrip$, then $\Ko_{abc}{}^d$ is $C^1$ there. In the space of vacuum solutions to Einstein's equations, there exists a neighborhood of Minkowski space-time in which this condition holds, obtained by a suitable choice of initial data \cite{pced}. If differentiability at $\scrip$ is weaker as, e.g., in \cite{dcsk} and \cite{bieri}, one has peeling only for $\Psi_4^\circ, \Psi_3^\circ$ and $\Psi_2^\circ$.  As noted above, the assumption in \emph{Definition 1} that $\go_{ab}$ is $C^4$ is made only to simplify the discussion; it is not essential for our results.

\subsection{The BMS group and its Poincar\'e reduction}
\label{s2.2}

We will first briefly recall the structure of the BMS group and then summarize a construction that leads to its \Poincare reduction in space-times in which the intrinsic connection $\Do$ on $\scrip$ tends to a ``classical vacuum" in the distant past, i.e., for which ${}^\star\!{\Ko}_{ab} \to 0$ as $\uo \to -\infty$. This additional condition is quite weak for isolated gravitating systems of astrophysical interest.

The BMS group $\BMS$ is the group of diffeomorphisms on $\scrip$ that preserves the universal structure it carries --i.e., the common structure it inherits from all space-times satisfying \emph{Definition 1}. From the summary presented in the  last subsection, it follows that this universal structure consists of the $\mathbb{S}^2 \times \mathbb{R}$ topology of $\scrip$, together with the collection of pairs of fields $(\qo_{ab}, \no^a)$ such that: (i) $\qo_{ab}$ is a degenerate metric of signature 0,+,+ with $\qo_{ab}n^{b} =0$ and $\mathcal{L}_{\no}\,\qo_{ab} =0$; \,  and, (ii) any two pairs $(\qo_{ab}, \no^a)$ and $(\qo^\prime_{ab}, {\no^\prime}^a)$ in the collection are related by a
conformal rescaling,
\be \label{rescaling} \qo^\prime_{ab} = \muo^2\, \qo_{ab} \quad {\rm and} \quad
{\no}^{\prime,\,a} = {\muo}^{-1} \no^a\, , \ee
where $\Lie_{\no}\muo = 0$. Note that, because 2-spheres carry a unique conformal structure, every $\qo_{ab}$  in this collection is conformal to a unit 2-sphere metric.

At the infinitesimal level, elements of the Lie algebra $\b$ of $\BMS$ can be naturally represented by vector fields $\xi^{a}$ on $\scrip$,\, motions along which preserve the universal structure, i.e., that satisfy:
\be \label{bms-xi} \Lie_{\xi} \qo_{ab} = 2\alphao\, \qo_{ab},\quad {\rm and} \quad
\Lie_{\xi} \no^{a} = - \alphao\, \no^{a} \ee
for some function $\alphao$ on $\scrip$ satisfying $\Lie_{\no}\, \alphao =0$. In particular, the vector fields $\xi^a = \fo n^a$ with $\Lie_{\no} \fo=0$ satisfy this condition, where $\fo$ is a function with \emph{conformal weight 1} (see the rescaling property of $\no^a$ in (\ref{rescaling})). Furthermore, the subspace $\s$ they form in the Lie algebra $\BMS$ of $\BMS$ is a Lie ideal in the sense that $[\xi, \, \fo \no^a] \in \s$ for all $\xi^a \in \b$ and all $\fo \no^a \in \s$. This is the Lie ideal of BMS \emph{supertranslations.} Next, Eq.(\ref{bms-xi}) implies that each BMS vector field $\xi^a$ can be unambiguously projected to a vector field $\bar{\xi}^a$ on the 2-sphere $\bar{S}$ of generators of $\scrip$ which is a conformal Killing field of the metric $\mathring{\bar{q}}_{ab}$ thereon. Because $\mathring{\bar{q}}_{ab}$ is conformal to the unit 2-sphere metric, it follows that the Lie algebra of the projected vector fields $\bar{\xi}^a$ is isomorphic to the Lorentz Lie algebra $\mathfrak{l}$. Therefore, $\BMS$ is a semi-direct sum of the Abelian Lie algebra of supertranslations and the Lorentz Lie algebra. Returning to finite diffeomorphisms, we conclude that $\BMS$ is the semi-direct product, $\BMS = \S\, \ltimes \L$, of the group $\S$ of supertranslations with the Lorentz group $\L$. Now, because the space of generators of $\scrip$ is topologically $\Sbb^2$, one can show that $\BMS$ also admits a unique normal, 4-dimensional Abelian subgroup $\mathcal{T}$ \cite{sachs2}. This is a subgroup of $\S$ and in Minkowski space-time it coincides with the group of space-time translations. Nonetheless, while the \Poincare group admits only a 4-parameter family of Lorentz subgroups, the BMS group admits an infinite parameter family, any two being related by a supertranslation.  In the standard treatments, this then leads to the supertranslation ambiguity in the definition of angular momentum.

A natural avenue to reduce this ambiguity is to extract a preferred \Poincare subgroup of the BMS group $\BMS$  
by imposing physically reasonable boundary conditions as one approaches $\inot$ along $\scrip$.  As before, let us denote by $\uo$ any affine parameter of $\no^a$\, so that as one approaches $\inot$ along $\scrip$, $\uo \to -\infty$. For definiteness and simplicity, in this paper we will specify fall-offs along $\scrip$ using $O(1/{\uo}^n)$ conditions and assume furthermore that if a field $F$ is $O(1/{\uo}^n)$\,\,---i.e., if $\uo^n F$ admits a finite $\uo\to -\infty$ limit---then $\big((\Lie_{\no})^m\, F\big)$ is $O(1/\uo^{m+n})$.% 
\footnote{\label{fn4} We use $O(1/{\uo}^n)$ fall-off because it is tailored to the `radial-angular' interplay in the notion of $C^{>1}$ differentiability imposed along space-like directions. It is possible to replace it with a weaker  
$1/{\uo}^{n-1 +\epsilon}$ fall-off. But the intermediate equations become less transparent because of factors of $\epsilon$.}
Then the additional condition is: 
\vskip0.1cm
\indent\emph{Condition 1: One restricts oneself to space-times $(\hat{M}, \hat{g}_{ab})$ satisfying \emph{Definition 1} in \\ \indent which, in addition, ${}^\star{\Ko}_{ab}$ is $O(1/\uo)$} as $\uo \to -\infty$ along $\scrip$.
\vskip0.1cm 
\noindent If Condition 1 is satisfied in any one divergence-free conformal frame, it is satisfied in all. The fall-off requirement is mild because ${}^\star{\Ko}_{ab}$ contains only the radiative information that is encoded in the connection $\Do$. In particular, Condition 1 is automatically satisfied if curvature becomes stationary as one approaches $\inot$ along $\scrip$ (see, e.g., \cite{ajetn,adlk-constraints}) as is generally assumed in the post-Newtonian literature on compact binary mergers \cite{lb-rev}. For the class of space-times satisfying Condition 1, one can introduce additional structures on $\scrip$. Then the requirement that these be preserved (in addition to the universal structure) reduces the BMS group $\BMS$ to a canonical \Poincare subgroup $\bmsP$. As we now summarize, this procedure can be carried out in two equivalent ways.  

The first way to select $\bmsP$ emphasizes connections $\{\Do \}$ \cite{aa-rad}.  Fix a divergence-free conformal frame $(\qo_{ab},\no^a)$ on $\scrip$ and consider derivative operators $\Do$ satisfying (\ref{Do}). We will denote by $\{\vac \}$ the  equivalence classes of connections for which ${}^\star\!{\Ko}_{ac}\, \=\,0$ (so that their news tensor $N_{ab}$ also vanishes identically \cite{abk1}). Let us make a small detour to note a key property of these $\{\vac \}$. They carry no dynamical information: their action $\{\vac_a\} K_b$ on any 1-form $K_b$ on $\scrip$ is completely determined by $\qo_{ab}$.% 
\footnote{Recall that $\{\vac_a\} K_b$ denotes the equivalence class of tensor fields $ \{ \vac_a K_b + f \qo_{ab}\, \no^c K_c \}$,\, where $\vac$ is any connection in $\{ \vac \}$ and $f$ an arbitrary function on $\scrip$.}
In this sense their curvature is trivial. Borrowing terminology from Yang-Mills theory, $\{\vac\}$ is referred to as a `classical vacuum'. Let us now recall the interplay between classical vacua and the action of the BMS group. Consider a BMS supertranslation $\xi^a$. Then in Eq.~(\ref{bms-xi}) we have $\alphao =0$, i.e., the action of $\xi^a$ preserves the conformal frame $(\qo_{ab}, \no^a)$. However, generically it does not preserve a given $\{ \vac \}$; it does so if and only if $\xi^a$ is a BMS translation. Passing from Lie algebras to the Lie groups, while $(\qo_{ab}, \no^a)$ is left invariant by the entire supertranslation group $\S$,\, any given classical vacuum $\{ \vac \}$ is left invariant only by its translation subgroup. Next, if pass to another divergence-free conformal frame via $\Omegao \to  \muo \Omegao$ in $M$,  we have 
\be \label{conformal1} \qo_{ab} \to \qo^{\prime}_{ab} = \muo^2\qo_{ab} \quad {\rm and} \quad \no^a \to \no^{\prime} = \muo^{-1} \no^a\, , \ee
on $\scrip$,\, while $\{ \vac \}$ transforms into the classical vacuum $\{\vac^{\prime} \}$ in the $(\qo^\prime_{an}, \, \no^{\prime\, a})$ conformal frame, given by 
\be \label{conformal2} \{\vac_a\} K_b \to \{\vac^{\prime}_a \} K_b =  \{\vac_a\} K_b - 2\muo^{-1} K_{(a} \vac_{b)}\muo \,.
\ee
Now, under a general BMS transformation, the conformal frame changes via (\ref{conformal1}) for some $\muo$. Therefore, it is natural to seek the subgroup of the BMS group which sends $\{\vac\}$ to $\{ \vac^\prime \}$ via (\ref{conformal2}). This is precisely a \Poincare subgroup $\bmsP$ of $\BMS$  (determined by the given vacuum $\{ \vac \}$). The space of all vacua is naturally isomorphic with the quotient $\S/\T$ of the BMS supertranslation group by its translation subgroup \cite{aa-rad}.

Let us now return to a physical space-time $(\hM, \hg_{ab})$ for which ${}^\star{\Ko}_{ab}$ vanishes in the distant past of $\scrip$, as in our Condition 1. Then, the $\{\Do\}$ induced on $\scrip$ by $\nabla$ in any divergence-free conformal completion tends to a unique classical vacuum $\{ \vac \}$ in the distant past. That is, there exists a $\{\vac\}$ such that for any (smooth and uniformly bounded) 1-form $V_a$ on $\scrip$, we have
\be \lim_{\uo \to -\infty} \big( \{ \Do_a\} - \{ \vac_a\} \big) V_b =0 \, , \ee
and under the conformal rescalings $\Omegao \to  \muo \Omegao$, the vacuum $\{ \vac\}$ transforms via (\ref{conformal2}). Therefore, if we demand the `past vacuum' should also be preserved, in addition to the universal structure, the asymptotic symmetry group of the given space-time $(\hM, \hg_{ab})$ reduces from $\BMS$ to  $\bmsP$ \cite{aa-rad,aa-yau}. By construction, this reduction is associated with the limit of structures on $\scrip$ as one approaches $\inot$; it is not a property of $\scrip$ by itself.

The second way to select this $\bmsP$ uses the extra structure associated with the Newman-Penrose formalism. Let us make a small detour to fix the notation used in that framework. One starts with the observation that one can always further restrict the conformal freedom on $\scrip$ by demanding that the metric on any of its 2-sphere cross-sections be the {unit, round 2-sphere metric}. This is always possible and such metrics $g_{ab}$ are said to constitute \emph{Bondi conformal frames}. In these frames, the vector field $\no^a$ represents a BMS \emph{time-translation} (rather than a general supertranslation). One fixes an affine parameter $\uo$ of $\no^a$ and introduces three other null vectors using the $\uo ={\rm const}$ cross-sections: $\ello^a$ which is normal to these cross-sections and $\mo^a, \bar{\mo}^a$ which are tangential to them, such that they constitute a null tetrad; i.e., the only non-zero scalar products are $\ello_a \no^a =-1$ and $\bar{\mo}_a \mo^a = 1$. 
The intrinsic metric on the $\uo = {\rm const}$ cross-sections is given by ${\mathring{\underbar{q}}}_{ab} = 2 \mo_{(a} \bar\mo_{b)}$ and the shear of these cross-section is given by $\sigmao^{ab} = \big({\mathring{\underbar{q}}}^{ac} {\mathring{\underbar{q}}}^{bd}\, - \f{1}{2} \mathring{\underbar{q}}^{ab} {\mathring{\underbar{q}}}^{cd}\big)\, \Do_c \ello_d$. The five components of ${}^\star{\Ko}_{ab}$ then correspond to two complex and one real Newman-Penrose scalars:
\be \Psi_4^\circ = \Ko_{abcd}\,\no^a \bar{\mo}^b \no^c \bar{\mo}^d,\quad \Psi_3^\circ = \Ko_{abcd}\,\no^a\bar{\mo}^b \no^c \ello^d,\quad {\rm and}\quad 2\, {\rm Im} \Psi_2^\circ = \Ko_{abcd}\,\no^a\ello^b \mo^c \bar{\mo}^d.\ee
Thanks to Einstein's equations (and Bianchi identities) that relate these Newman-Penrose scalars to the shear and its derivatives, our Condition 1 on ${}^\star\Ko_{ab}$ implies that the shear $\sigmao = \sigmao_{ab} {\mo}^a {\mo}^b$ of the $\uo = {\rm const}$ foliation has asymptotic behavior $\sigmao (\uo, \theta, \phi) = \sigmao^{(0)}(\theta,\phi) + \f{1}{\uo}\, \sigma^{(1)}(\theta,\phi) + \ldots$, where the leading term $\sigma^{(0)}(\theta,\phi)$ is ``purely electric", i.e. of the form $\mo^a \,\mo^b\, \Do_a \Do_b f$ (or, $\eth^2 f$) for a real function $f$ \cite{etnrp,etnrp2,adlk-constraints}. Einstein's equations and the Bianchi identities then imply that the Bondi News $N_{ab}$ and $\Psi_3^\circ$ fall off as $1/\uo^2$, and $\Psi_4^\circ$ as $1/\uo^3$. (These fall off conditions ensure the total fluxes of BMS momenta from $\uo = -\infty$ until any cross-section $C$ of $\scrip$ are all finite \cite{aams}.)

With these preliminaries out of the way, we can summarize the second way  \cite{etnrp,adlkJ} to extract a preferred \Poincare group of the BMS group $\BMS$. Since the asymptotic shear $\sigma^{(\circ)}(\theta,\phi)$ is purely electric, it can be transformed away by applying a super-translation to the initial family $\uo ={\rm const}$ of cross-sections. The supertranslation does not change the conformal frame $(\qo_{ab},\, \no^a)$ at $\scrip$; it just yields another affine parameter $\uo^\prime$ of $\no^a$. The resulting family ${\uo}^\prime = {\rm const}$ cross-sections has the property that their shear vanishes in the limit $\uo \to -\infty$. The family is not unique, but the only freedom is that of making
a BMS translation which results in another family with the same property. Thus, if ${}^\star{\Ko}_{ab}$ is $O(1/\uo)$ on $\scrip$, one obtains a preferred 4-parameter family of  cross-sections on $\scrip$ completely characterized by the property that their shear has the asymptotic form $\sigmao (\uo, \theta, \phi) = \f{1}{\uo}\, \sigma^{(1)}(\theta,\phi) \,+ \ldots$ as $u\to -\infty$. The subgroup of the BMS group $\BMS$ that preserves this family is again a \Poincare subgroup. In fact, it is precisely the $\bmsP$\, obtained using the preferred classical vacuum ${\Do}_\circ$ (see Appendix A of \cite{aa-rad}).\\

\emph{Remarks:}\smallskip

1. Let us begin by listing three simplifications that occur in Bondi frames which will be used in section \ref{s3} as well as in the accompanying paper \cite{ak-J}.\\
\indent (i) A BMS supertranslation $\xi^a = \fo\, \no^a$ is a translation if and only if\,\, $\Do_a \Do_b \fo  \propto\, \qo_{ab}$.\,\, This is equivalent to asking that $\fo$ be a linear combination of the first four spherical harmonics, $\fo = \fo_0\,Y_{0,0}(\theta,\phi) +\sum_{m=-1}^1\, \fo_m\, Y_{1,m}(\theta,\phi)$ for some real constants $\fo_0, \fo_m $.\\
\indent (ii) In a Bondi conformal frame, the $Y_{1,m}$ also feature in the expression of generators of the Lorentz subgroup that leaves a $\uo = {\rm const}$ cross-section invariant. On the cross-section, the generators of rotations are linear combinations $\zeta^a = \sum_{m=-1}^{m=1} C_m\, \epsilono^{ab} \Do_b Y_{1,m}$ and boosts are linear combinations ${\underline{\zeta}}^a = \sum_{m=-1}^{m=1} \underbar{C}_m\, \qo^{ab} \Do_b Y_{1,m}$, where $C_m$ and $\underbar{C}_m$ are real constants and $\epsilono^{ab},\,\, \qo_{ab}$ the alternating tensor and the metric on the cross-section.\\
\indent (iii) Finally, in a Bondi conformal frame, the News tensor $\No_{ab}$ is just the trace-free part, $\rm{TF}\, (\mathring{\underbar{S}}_{ab})$, of the pull-back of the Schouten tensor $\mathring{S}_{a b}$ to $\scrip$. \vskip0.1cm

 2. The shear-free cross-sections of the second method can be extracted from the classical vacuum $\{ \vac \}$ of the first method as follows. Let $(\qo_{ab}, \no^a)$ be a Bondi conformal frame. The action $\Do_a K_b$ of any $\Do$ on `horizontal' ${K}_b$ is universal, determined by $\qo_{ab}$ (where,  $K_b$ is horizontal if it is a lift to $\scrip$ of  1-forms $\bar{K}_a$ on the 2-sphere of the integral curves of $\no^a$). Thus, $\Do$ is completely determined by its action on any one `transversal' 1-form $\ello_b$ satisfying $\no^a \ello_a =-1$. Therefore, one might expect that this action should be trivial in an appropriate sense for vacuum connections $\{ \vac \}$. This is indeed the case: Given any vacuum $\{ \vac \}$, there is a unique $\vac \in \{ \vac \}$ and an associated (unique) 1-parameter family of cross-sections, $\uo = {\rm const}$, such that $\lo_b = -\vac_b \uo$ satisfies $\vac_a \lo_b =0$ (where, as usual, $\uo$ is an affine parameter of $\no^a$). Hence, these cross-sections are shear-free w.r.t. $\vac$. Next, since the shear $\sigmao_{ab}$ of any cross-section is trace-free, it then follows that the shear of any cross-section $\uo = {\rm const}$ vanishes w.r.t. any other connection in $\{\vac\}$. Furthermore, if we shift this cross-section by a BMS translation\, $\uo \to \uo^\prime = \uo + f_0 + \sum_m f_m  Y_{1,m}(\theta, \varphi)$, then the shear of the new family of cross-sections again vanishes (because $\vac_a \vac_b Y_{1,m} \propto \qo_{ab}$). Finally, if we make a conformal transformation to a general divergence-free frame, $\sigmao_{ab}$ just rescales by the conformal factor and hence continues to vanish. Thus, any given vacuum $\{ \vac \}$ determines a 4-parameter family of cross-sections of $\scrip$ that are related by BMS translations which are shear-free in any divergence-free frame.%
\footnote{However, in a non-Bondi conformal frame, the vector field $\no^a$ is not a BMS translation. Therefore, the preferred 4-parameter family of cross-sections is not left invariant by the action of $\no^a$. Reciprocally, in a non-Bondi frame, the 1-parameter family of cross-sections $\uo ={\rm const}$ does not belong to the preferred family for any choice of affine parameter $\uo$ of $\no^a$.}
Therefore, the subgroup $\bmsP$ of $\BMS$ that leaves the given vacuum invariant also leaves the 4-parameter family of shear-free cross-sections invariant and vice-versa. To summarize, the \Poincare subgroups of $\BMS$ selected by the two methods agree. (For details, see Appendix A in \cite{aa-rad}.) 

3. Note however, that one can also carry out a similar reduction using $i^+$ in place of $\inot$ and\, $\bmsP$\, and \,$\mathfrak{p}^{\rm bms}_{i^+}$\, are generically distinct \Poincare subgroups of the BMS group; they agree if and only of the total gravitational memory vanishes. This is why $\bmsP$ is not routinely used in the discussions involving asymptotic flatness only at null infinity. Nonetheless,\, $\bmsP$ \, is the one that features in the comparison of structures at null and spatial infinity. For comparison with timelike infinity, such as in \cite{cgw}, using the $\mathfrak{p}^{\rm bms}_{i^+}$ would be more appropriate. See \cite{adlkJ} for how the two groups and their charges are related at null infinity. However, as we discuss in section \ref{s5}, this does \emph{not} mean that we can abandon the BMS group on $\scri$ and work just with $\bmsP$ or $\mathfrak{p}^{\rm bms}_{i^+}$ in the study of gravitational waves; both $\B$ and $\bmsP$ are important to the discussion of isolated systems, and the context, i.e., the issue under discussion decides which is the relevant one.\vskip0.1cm

\subsection{Spatial infinity $\inot$}
\label{s2.3}

We will now recall the structure of the gravitational field at spatial infinity. Let us begin with the definition of asymptotic flatness at spatial infinity tailored to the 4-dimensional setting without a 3+1 split, introduced in \cite{aa-ein}.\\ 
{\textbf{Definition 2}} A physical space-time $(\hat{M}, \hat{g}_{ab})$ is said to be \emph{asymptotically flat at spatial infinity} if there exists a space-time $(M, g_{ab})$ with a preferred point $\inot$, and an embedding of $\hat{M}$ into $M$ with $g_{ab} = \Omega^2 \hat{g}_{ab}$ on $\hat{M}$ such that:

i) $\bar{J}(\inot)= M \setminus \hat{M}$, where $\bar{J}(\inot)$ is the closure of the region in $M$ causally related to $\inot$;

ii) $\Omega$ and $g_{ab}$ are $C^3$ on $M\setminus \inot$, while $g_{ab}$ is $C^{>0}$ at $\inot$, where $\Omega =0$\,\, $\nabla_a \Omega =0$\,\, and \indent\indent $\nabla_a\nabla_b \Omega = 2\, g_{ab}$; 

iii)\, $\hat{g}_{ab}$ satisfies Einstein's equations $\hat{R}_{ab} - \frac{1}{2} \hat{R} \hat{g}_{ab}  = 0$ in a neighborhood of $\inot$.\\ 
The point $\inot$ represents spatial infinity because in the conformal completion it is space-like w.r.t. \emph{all} points on the physical space-time manifold $\hat{M}$. In section \ref{s3} we will see that the null cones of $\inot$ serve as $\scripm$. However, in this section we will focus \emph{only on asymptotic flatness in space-like directions}.\medskip

The role of these conditions is as follows. Condition i) requires that the set of points in $M$ that are space-like related to $\inot$ is precisely the physical space-time manifold $\hat{M}$; thus $\inot$ represents the point at spatial infinity of the physical space-time. Condition on $\Omega$ in ii) ensure that the conformal factor $\Omega$ falls off as $\sim ``1/{r}^2"$, just as it does at $\inot$ of Minkowski space-time. (In Minkowski space, in terms of coordinates $\hat{u},\hat{v},\hat{\theta},\hat{\phi}$,\, the conformal factor  $\Omega$ falls off as $1/\hat{u}\hat{v}$. As one approaches $\inot$ along space-like lines, we have $\Omega \sim 1/\hat{r}^2$ while, since $\scripm$ are approached along $\hat{u}= {\rm const}$ and $\hat{v}={\rm const}$ surfaces respectively, we have $\Omega \sim 1/\hat{v} \sim 1/\hat{r}$ as one approaches $\scrip$ and  $\Omega \sim 1/\hat{u} \sim 1/\hat{r}$ as one approaches $\scrim$.) The $C^3$ differentiability on $M \setminus \inot$ enables us to use Bianchi identities in ${M}$ while exploring the implications of field equations in the limit to $\inot$.%
\footnote{In \emph{Definition 1} we required $g_{ab}$ to be $C^4$ because in the discussion of balance laws of BMS momenta one has to use the Bianchi identity on $K_{abcd} = \lim \Omega^{-1} C_{abcd}$ at $\scrip$. In the discussion of $\inot$ $C^3$ differentiability suffices because one only needs the Bianchi identity on $C_{abcd}$.} 
Finally, condition iii) can be weakened to allow for matter fields in a neighborhood of $\inot$ so long as the physical stress-energy tensor $\hat{T}_{ab}$ admits a (possibly) direction dependent limit to $\inot$. 

The $C^{>0}$ differentiability of $g_{ab}$ at $\inot$ is rather peculiar, and this subtlety was responsible for a substantial delay in the exploration of the structure of $\inot$. The awkward differentiability is unavoidable because the ADM mass is encoded in the \emph{radial discontinuities} of the connection $\nabla$ at $\inot$ \cite{aa-padova}; it vanishes if ${g}_{ab}$ is $C^1$ at $\inot$, and ceases to be well-defined if $g_{ab}$ is only\, $C^0$\, there! The notion of  $C^{>0}$ differentiability is introduced in detail in Appendix A of \cite{aarh}. We will only summarize the properties that we will need. In essence, a metric $g_{ab}$ is $C^{>0}$ if it is continuous and, in addition, the connection $\nabla$\,---or the Christoffel symbols in appropriate (i.e. $C^{>1}$) charts---have only finite radial discontinuities at $\inot$. They admit direction dependent limits at $\inot$ which are \emph{regular} in the sense that: (i) the limits are smooth with respect to the direction of approach, i.e. w.r.t. the (hyperboloidal) angles defined by the directions of approach to \,$\inot$;\, and, (ii) the operation of taking the \emph{angular} derivatives and limit to $\inot$ commute.  Given a physical space-time $(\hat{M}, \hat{g}_{ab})$ satisfying \emph{Definition 2}, it is natural to consider space-like hypersurfaces $\Sigma$ in $M$ that pass through $\inot$ and are $C^{>1}$ there. Then the initial data induced by the physical metric $\hat{g}_{ab}$ on  $\Sigma \setminus \inot$ satisfy the standard ADM fall off conditions (see Appendix A in \cite{am3+1}). 

To spell out the key implications of `regular direction dependent limits' let us consider the approach to $\inot$ along curves to which $\eta^a := \nabla^a \Omega^{\f{1}{2}}$ is tangential. Thanks to the conditions of \emph{Definition 2}, we have $\eta^a \eta_a \=1$ at $\inot$. If a tensor field $T^{a\ldots b}{}_{c\ldots d}$ admits a regular direction dependent limit at $\inot$, the limit $\mathbf{T^{a\ldots b}{}_{c\ldots d}}(\eta)$ can be regarded as a field on the hyperboloid $\H$ of unit space-like vectors $\eta$ in the tangent space $T_{\inot}$ of $\inot$. In general this field will have components tangential as well as orthogonal to $\H$ and their projections provide a set of fields defined intrinsically on $\H$. Let us denote the intrinsic metric on $\H$ by $\bfh_{ab}(\eta) = \lim_{\to \inot} (g_{ab} - \eta_a \eta_b)$. {If the limit $\mathbf{T^{a\ldots b}{}_{c\ldots d}}(\eta)$ of a $C^{>0}$ field is tangential to $\H$ then $\lim_{\to\inot}\, \Omega^{\f{1}{2}}\,\nabla_m T^{a\ldots b}{}_{c\ldots d} = \mathbf{ D_m\,T^{a\ldots b}{}_{c\ldots d}}(\eta)$, where $\bfD$ is the derivative operator on $\H$ compatible with $\bfh_{ab}$.}%
\footnote{This is the precise meaning of condition ii) above on commutativity of the operations of taking angular derivatives and limits to $\inot$ because the left side provides the `angular derivatives in a Cartesian chart' around $\inot$, before taking the limit.}

Note that there is freedom in the choice of the conformal factor: $\Omega \to \Omega^\prime = \omega \,\Omega$ provides another permissible conformal factor if $\omega$ is $C^{>0}$ at $\inot$ and  $C^3$ elsewhere, and satisfies $\omega\, \=\,1$ at $\inot$. Since $\nabla_a \omega$ admits regular direction-dependent limits to $\inot$, it follows that $\omega = 1 + \Omega^{\f{1}{2}} {\alpha}$, where ${\alpha}$ admits a regular direction-dependent limit ${\bfalpha}(\eta)$ at $\inot$. Thus, all metrics $g_{ab}$ agree \emph{at} $\inot$ and their leading order difference is registered in the function ${\bfalpha}(\eta)$ on $\H$. Two metrics $g_{ab}$ and $g^\prime_{ab}$ are said to be equivalent if the relative ${\bfalpha}(\eta)$ vanishes. Each equivalence class is referred to as a \emph{ripple} on the asymptotic metric. Now, derivative operators $\nabla^\prime_a$ and $\nabla_a$ compatible with $g_{ab}$ and $g^\prime_{ab}$ are related by
\be \label{conformal3}  \nabla^\prime_a  K_b = \nabla_a K_b  - 2\omega^{-1} (\delta_{(a}^c\,\nabla_{b)}\omega - \nabla^c \omega\, g_{b})\, K_c \, \ee
% I just put it in the same form as (2.9)
and since $\omega = 1 + \Omega^{\f{1}{2}} {\alpha}$, we have
\be \lim_{\to\inot} \nabla_a \omega = {{\bfalpha}}(\eta)\, \eta_a + \mathbf{D}_a {\bfalpha}(\eta)\, . \ee
Therefore, $\nabla^\prime_a$ and $\nabla_a$ agree asymptotically  if and only if ${\bfalpha}(\eta)\=0$, i.e., $g^\prime_{ab}$ and $g_{ab}$ belong to the same ripple. \emph{Using just the available asymptotic structure one cannot distinguish between two metrics in the same ripple.} The point\, $\inot$,\, the metric $g_{ab}$ there, and the collection of associated ripples constitutes the universal structure at spatial infinity  --the structure that is common to all space-times that are asymptotically flat at spatial infinity in the sense of \emph{Definition 2} \cite{aa-ein}. As discussed below, the asymptotic symmetry group $\Spi$ at spatial infinity preserves this structure.

Next, the asymptotic curvature of $g_{ab}$ also provides us with asymptotic fields that carry invariantly defined information about physical properties of the isolated system under consideration, and thus vary from one space-time to another. Since $g_{ab}$ is $C^{>0}$, it follows that $\Omega^{\f{1}{2}}\, R_{abcd}$ admits a regular direction dependent limit $\mathbf{R}_{abcd}(\eta)$ at $\inot$. The asymptotic Weyl curvature is conveniently encoded in two smooth fields tangential to $\H$,
\be \label{EandB} \mathbf{E}_{ac} := \mathbf{C}_{abcd}(\eta)\, \eta^b\, \eta^d \qquad {\rm and}\qquad 
\mathbf{B}_{ac} := \mathbf{{}^\star C}_{abcd}(\eta)\, \eta^b\, \eta^d \, ,\ee
defined intrinsically on $\H$. They can be thought of as ``electric and magnetic parts" of the asymptotic Weyl curvature, but now defined using space-like unit vectors $\eta^a$ at $\inot$ rather than unit time-like normals to a Cauchy surface. The Bianchi identity satisfied by the curvature of the physical metric $\hat{g}_{ab}$ implies that they satisfy the following field equations:
\be \label{spieqs} \bfD_{[a}\bfE_{b]c}\, =\, 0 \qquad {\rm and}\qquad \bfD_{[a}\bfB_{b]c}\, =\,0\, .\ee
As at $\scri$, the Ricci part --or rather the Schouten tensor $S_{ab}$-- of the asymptotic curvature of $g_{ab}$ provides potentials $\bfE = \bfS_{ab} \eta^a \eta^b$  and $\bfK_{ab} = \bfh_a{}^c\, \bfh_b{}^d\,  \bfS_{cd} - \bfE\, \bfh_{ab}$ for the asymptotic Weyl curvature through the Bianchi identity. We have
\be \bfE_{ab} = \bfD_a \bfD_b \bfE + \bfE\, \bfh_{ab} \qquad {\rm and}\qquad  \bfB_{ab} = - \textstyle{\f{1}{4}}\,\bfepsilon_{mna}\bfD^m \bfK^n{}_b \ee
where, as explained above, $\bfh_{ab}$ is the intrinsic metric on $\H$,\, and $\bfD$\,and\, $\bfepsilon_{abc}$ are the derivative operator and the alternating tensor $\bfh_{ab}$ defines on $\H$. These potentials will be useful in the \Poincare  reduction of the asymptotic symmetry group $\Spi$. (Note that the potential $\bfK_{ab}$ is unrelated to the asymptotic Weyl curvature ${}^\star\!\Ko_{ab}$ of sections \ref{s2.1} and \ref{s2.2}.)\\

\emph{Remarks:}

1. The Kerr metric satisfies the asymptotic conditions of \emph{Definition 2} \cite{aarh,herberthson}. The boost theorem of \cite{ycbdc,dcno} implies that one can specify vacuum initial data on a Cauchy surface such that the asymptotic conditions are preserved on surfaces obtained by any pre-specified finite boost. In these space-times, the asymptotic conditions at $\inot$ are satisfied on large but finite `wedges'. There is also a physical space-time description of our asymptotic conditions without reference to a completion \cite{rbbs} that supports the existence of a large class of solutions to Einstein's equations satisfying \emph{Definition 2.} Similarly, in terms of conformal completions, results of \cite{kroon1} also indicate that there is a large class of examples.

2. Given a conformal completion satisfying \emph{Definition 2}, there exists a 4-parameter family of other completions in which the differential structure at $\inot$ is inequivalent to that used in the first completion \cite{aa-log,pc-log}. This freedom corresponds to that of performing logarithmic translations in the physical space-time, first pointed our by Bergmann \cite{pgb}. It turns out that this inequivalence is physically harmless in the sense that it does not change the leading order physical fields $\bfE_{ab}$ and $\bfB_{ab}$ and the conserved quantities associated with the given isolated system. They also leave the potential $\bfK_{ab}$ of $\bfB_{ab}$ untouched \cite{aa-padova} but transform the potential $\bfE$ of $\bfE_{ab}$. As a consequence, this freedom can be eliminated by requiring an additional condition on the permissible conformal completion: demand that the function $\bfE$ on $\H$ be reflection symmetric \cite{aa-log}. We will make this assumption just to simplify the discussion. 

\subsection{The Spi group and its Poincar\'e reduction}
\label{s2.4}

As we noted above, the universal structure at spatial infinity consists of the point $\inot$, the metric $g_{ab}$ there, and the collection of \emph{ripples} on this asymptotic metric, encoded in the compatible connections $\nabla_a$ (the difference between any two being captured in the function $\bfalpha(\eta)$ on $\H$). The asymptotic symmetry group $\Spi$ at $\inot$ is then the quotient ${\rm Diff}/{\rm Diff}_\circ$ of the group ${\rm Diff}$ of diffeomorphisms on $M$ that preserves this structure by the subgroup ${\rm Diff}_\circ$ that leaves it untouched. We will now summarize its structure which was spelled out in \cite{aa-ein}. At the infinitesimal level, generators $\xi^a$ of ${\rm Diff}$ are vector fields that are $C^{>0}$ at $\inot$ and $C^3$ elsewhere. Since symmetries must leave $\inot$ and the metric at $\inot$ invariant and preserve the set of ripples, $\xi^a$ must satisfy the following equations at $\inot$:
\be \label{spisym} \xi^a\, \= \, 0; \qquad \lim_{\to \inot} \nabla_{(a} \xi_{b)} =0; \qquad {\rm and} \qquad \lim_{\to \inot} \big(\nabla_{a} \nabla_{(b} \xi_{c)}\, -\, 2 (\nabla_a \phi)\, g_{bc}\big) =0;\,  \ee
for some $C^{>0}$ function $\phi$ that vanishes at $\inot$. While a specific metric $g_{ab}$ has been used in (\ref{spisym}), these conditions on $\xi^a$ are independent of this choice.  Since $\xi^a$ is $C^{>0}$ at $\inot$ and  vanishes there, it has the form $\xi^a = \Omega^{\f{1}{2}}\, \zeta^a$ where $\zeta^a$ admits a regular direction dependent limit $\zeta(\eta)$ at $\inot$. The condition $\lim_{\to \inot} \nabla_{(a} \xi_{b)} =0$ implies only that $\zeta^a(\eta)$ is tangential to $\H$ and a Killing field of the metric $\bfh_{ab}$ thereon. Since $\H$ is the unit hyperboloid in $T_{\inot}$, it follows that the Lie algebra generated by these $\zeta^a$ is the Lorentz Lie algebra $\mathfrak{l}$.

If $\bfzeta^a(\eta) = 0$, i.e., if $\lim_{\to\inot} \Omega^{-\f{1}{2}}\, \xi^a$ vanishes, the diffeomorphism generated by $\xi^a$ leaves not only $\inot$ but also the tangent space $T_{\inot}$ invariant and only reshuffles the ripples, changing $\nabla_a$ at $\inot$ via the infinitesimal version of (\ref{conformal3}), with $\phi$ in (\ref{spisym}) serving as the infinitesimal change in the conformal factor $\omega$. These symmetry vector fields $\xi^a$ represent spi-supertranslations, each characterized by a direction-dependent vector $\bfphi_a(\eta) := \lim_{\to\inot} \nabla_a\phi$, or equivalently, by the smooth function $\bff(\eta) : = \eta^a \bfphi_a(\eta)$ on $\H$. They constitute an infinite dimensional Abelian subalgebra $\s_{\inot}$ of the spi-symmetry algebra $\g$. The 4-dimensional subspace $\mathfrak{t}_{\inot}$ of $\s_{\inot}$ spanned by $\bff(\eta)$ satisfying
\be \label{spitran} \big(\bfD_a \bfD_b  + \bfh_{ab} \big)\,\bff(\eta) =0\,  \ee
constitutes the \emph{translation} sub-Lie algebra of $\s_{\inot}$. In Minkowski space-time, this is precisely the standard translation subgroup and, in stationary space-times satisfying \emph{Definition 2}, the time-translation Killing field belongs to $\mathfrak{t}_{\inot}$ \cite{am3+1}. The asymptotic translations $\xi^a$ are also characterized by the fact that $\bfphi_a$ is a \emph{direction independent} (i.e. continuous) 1-form at $\inot$. Note that solutions $\bff$ to (\ref{spitran}) are linear combinations of the first four ``hyperboloid harmonics", defined using the Laplacian of the 3-metric $\bfh_{ab}$ (which is of constant curvature). Thus, there is close analogy with the characterization of the translation sub Lie algebra of the BMS Lie algebra at $\scrip$ (in Bondi conformal frames). Finally, as in the BMS case, the Spi Lie algebra $\mathfrak{g}$ is a semi-direct sum of the supertranslation Lie algebra $\s_{\inot}$ and the Lorentz Lie algebra $\mathfrak{l}$. In terms of groups, then, $\Spi = \S \ltimes \L$ is the semi-direct product of the group $\s_{\inot}$ of supertranslations with the Lorentz group. Thus, in their structure, $\BMS$ and $\Spi$ are very similar; the main difference is that while BMS supertranslations are labelled by functions $f(\theta,\,\varphi)$ of conformal weight 1 on a 2-sphere, Spi supertranslations are labelled by function $\bff(\eta)$ on the 3-dimensional hyperboloid $\H$.

As shown in \cite{aarh}, one can carry out a \Poincare reduction of the Spi group $\G$ by a well-motivated strengthening of the boundary conditions. In retrospect, the strategy is closely related to the one used for the \Poincare reduction of $\BMS$ in section \ref{s2.2}. The additional condition is again imposed on the ``magnetic part" of the asymptotic Weyl curvature, which is now encoded in $\bfB_{ab}$:\vskip0.1cm 
\indent 
{\emph{Condition 2: One restricts oneself to space-times $(\hat{M}, \hat{g}_{ab})$ satisfying \emph{Definition 1} in \\
\indent which  $\bfB_{ab} =0$, where, as in Eq.~\eqref{EandB}, $\bfB_{ab} = \lim_{\to\inot} \Omega^{\f{1}{2}}\, \eta^c\eta^d\, {}^\star C_{acbd}$.}}\vskip0.1cm
\noindent If this condition is satisfied in one completion satisfying \emph{Definition 2}, it is also satisfied in any other. The additional condition on $\bfB_{ab}$ is quite natural from several independent considerations. First, it is automatically satisfied if the physical space-time $(\hat{M}, \hat{g}_{ab})$ is either stationary or axisymmetric \cite{am-conserved}. Second, it is used crucially in Refs. \cite{rbbs,beig} to solve vacuum Einstein's equations near spatial infinity, order by order in powers of $1/\rho$ of a radial coordinate in the physical space-time  (that corresponds to our $\Omega^{\f{1}{2}}$). Modulo the important issue of convergence of the series, this analysis establishes the existence of a large class of asymptotically flat solutions without any symmetry in which the condition is satisfied. Third, $\bfB_{ab} =0$ is precisely the condition needed in twistor theory to introduce a global twistor space at spatial infinity, which can then be used to explore the total energy-momentum and angular momentum of a gravitating system \cite{shaw-twistor}. Finally, this condition is also used at $\scri$ of asymptotically anti-de Sitter space-time to extract anti-de Sitter group as the asymptotic group of symmetries  and the corresponding definition of anti-de Sitter charges \cite{am-ads,ad-ads} resolves a tension (concerning the first law of black hole mechanics) related to the definition of angular momentum of Kerr-anti de Sitter solutions \cite{gpp}. 

For the subclass of space-times that satisfies Condition 2, one can restrict the conformal freedom by imposing a new asymptotic restriction, and the requirement that it also be preserved (in addition to the universal structure) reduces the Spi group $\Spi$ to $\spiP$. One begins with the observation that when Condition 2 is satisfied, one can always choose a conformal metric $g_{ab}$ satisfying \emph{Definition 2} for which the tensor potential of $\bfB_{ab}$ vanishes, i.e., $\bfK_{ab} =0$ on $\H$ \cite{aarh}. For brevity, let us call a conformal metric ${g}_{ab}$ \emph{admissible} if it satisfies $\bfK_{ab} =0$. Now, given any two completions $g_{ab}$ and  $g^\prime_{ab} = \omega^2\, g_{ab}$ satisfying \emph{Definition 2}, their potential $\bfK_{ab}$ and $\bfK^\prime_{ab}$ can both vanish \emph{if and only if $\omega$ is $C^1$ at $\inot$}. (Recall that for two generic metrics satisfying only \emph{Definition 2}, the relative conformal factor is only $C^{>0}$). Thus, the restriction $\bfK_{ab} =0$ selects a subfamily of conformal metrics $g_{ab}$ each of which is only $C^{>0}$ at $\inot$ but for which \emph{the relative conformal factor} $\omega$ is in fact $C^1$. That is, it selects a specific 4-parameter family of ripples, any two being related by $\nabla \omega\mid_{\inot}$.

What happens under the action of an infinitesimal supertranslation $\xi^a$? $g_{ab}$ undergoes an infinitesimal conformal rescaling by $\phi$ which is, in general, direction-\emph{dependent} whence $K_{ab}$ of the rescaled metric no longer vanishes. It is only when $\xi^a$ is a translation that $\phi$ is $C^1$ at $\inot$ and the condition $K_{ab} =0$ is preserved. Consequently, the diffeomorphism generated by $\xi^a$ preserves the preferred 4-family of ripples if and only if this supertranslation is a translation. Returning to the full Spi group $\Spi$, it then follows that the preferred subfamily of ripples is preserved precisely by a single \Poincare group $\spiP$ of $\Spi$.

\emph{Remarks:}

1. A detailed examination reveals that the above \Poincare reduction can be achieved under a slightly weaker condition than $\bfK_{ab} =0$: It suffices to demand that ${\rm TF}\,\bfK_{ab} \equiv {\rm TF}\,\bfh_a{}^m\,\bfh_b{}^n\,\bfS_{mn} =0$, where TF stands for `trace-free part of'. This weaker condition already selects the preferred class of $g_{ab}$ related by a $C^1$ conformal factor $\omega$. We will use this fact in Section \ref{s3}\vskip0.1cm

2. Recall that ripples on the asymptotic metric are in 1-1 correspondence with the \emph{asymptotic} connections $\nabla$ at $\inot$. The group $\S_{\inot}$ of spi-supertranslations  acts on the ripples---and hence on the space of asymptotic connections---simply and transitively. Let us denote by $\{\nabla_{\inot}\}$ the equivalence classes of asymptotic connections selected by admissible metrics $g_{ab}$. Thus, for two connections in $\{\nabla_{\inot}\}$, the $\nabla_a \omega$ that relates them via Eq.(\ref{conformal3}) is $C^0$ at $\inot$. Each of these equivalence classes is (trivially) left invariant by spi-translations and the quotient $\S_{\inot}/\mathfrak{T}_{\inot}$ acts on the space of equivalence classes $\{\nabla_{\inot}\}$ simply and transitively. \vskip0.1cm

3. Thus, the situation is completely analogous to that at $\scrip$. The equivalence classes $\{\nabla_{\inot}\}$ 
are analogous to the `classical vacua' $\{\Do\}$ on $\scrip$; under allowable conformal rescalings, they transform via (\ref{conformal3}) and (\ref{conformal2}) respectively. The group $\mathcal{T}$ of $\BMS$ translations leaves each $\{\vac\}$ invariant and the quotient $\S/\mathcal{T}$ acts simply and transitively on the space of classical vacua. If $\lim_{\uo \to -\infty}\, {}^\star{K}_{ab}$ vanishes on $\scrip$, $\{\vac\}$ on $\scrip$ tends to a classical vacuum $\{\Do_{\circ}\}$ and the \Poincare subgroup $\bmsP$ is obtained by restricting ourselves to the subgroup of $\BMS$ that preserves this $\{\vac\}$. Similarly, at $\inot$, if the magnetic part $\bfB_{ab}(\eta)$ of the asymptotic Weyl curvature vanishes, we can select a preferred equivalence class of ripples $\{\nabla_{\inot}\}$ and the \Poincare group $\spiP$ is obtained by restricting ourselves to the subgroup of the Spi group $\Spi$ that preserves this $\{\nabla_{\inot}\}$.

\section{Matching of structures at $\scri$ and $\inot$}
\label{s3}

This section is divided into two parts. In the first we discuss the relation between conformal completions that provide divergence-free frames at $\scrip$, and those in which $\inot$ serves as the vertex of the light cone $\scri$. In the second we present the definition of \emph{AM} space-times by strengthening \emph{Definition 2} through `gluing conditions' on the behavior of geometric fields as one approaches $\inot$ along space-like directions and along $\scrip$.

\subsection{Conformal frames}
\label{s3.1}

In section 2, we considered asymptotic flatness in null and spatial directions separately. In this section we wish to consider space-times that are asymptotically flat in both regimes in a manner that ties $\scripm$ to $\inot$ appropriately. Let us begin by recalling (essentially from \cite{aa-ein}) the notion of space-times that are asymptotically empty and flat at null and spatial infinity (AEFANSI).

{\textbf{Definition 3:}} A physical space-time $(\hat{M}, \hat{g}_{ab})$ is said to be {AEFANSI} if there exists a conformal completion $(M, g_{ab})$ with a preferred point $\inot$, and an embedding of $\hat{M}$ into $M$, such that $g_{ab} = \Omega^2 \hat{g}_{ab}$ on $\hat{M}$; and, \vskip0.1cm 
i) $\bar{J}(\inot)= M \setminus \hat{M}$, where $\bar{J}(\inot)$ is the closure of the region in $M$ causally related to $\inot$;\\
\indent ii) $\Omega$ and $g_{ab}$ are $C^4$ on $M \setminus \inot$ and $C^2$ and $C^{>0}$, respectively, at $\inot$. On the boundary \indent $\dot{J}(\inot)$ of $\bar{J}(\inot)$ in $M$ they satisfy:\\ 
\indent (a) $\Omega \=0$;\, (b) on $(\dot{J}\,\setminus\, \inot)$,\, $\nabla_a \Omega \not= \,0$; and,\, (c) at $\inot$,\,\,$\nabla_a\Omega\, \=\,0$,\, $\nabla_a\nabla_b \Omega\, \= \,2 g_{ab}$;\, and,\\
\indent iii) In a neighborhood $\mathcal{N}$ of $\dot{J}(\inot)$, $M$ is strongly causal and time orientable in $N$ and \indent the physical metric $\hat{g}_{ab}$ satisfies Einstein's vacuum equations in $\mathcal{N}\cap \hat{M}$.\medskip   

The causality and time-orientability requirements imply that $(\dot{J}(\inot)\setminus \inot)$ has two disconnected components, one to the future of $\inot$ and the other to the past, that serve as $\scripm$ of the physical space-time. Thus, null infinity is introduced here in a way that is quite different from that in the Bondi-Sachs \cite{bondi,sachs} or Penrose \cite{rp} frameworks: There is no reference to the outgoing null surfaces $u= {\rm const}$ in the physical space-time introduced in the Bondi-Sachs expansions of the physical metric,\, nor to the null geodesics used in Penrose's notion of asymptotic simplicity.  Nonetheless, rather surprisingly, $\scripm$ has all the properties that are normally used in the analysis of gravitational waves, including those summarized in section \ref{s2.1}. In particular, if we restrict ourselves only to the part $\hat{M}\, \cup\, (\dot{J}^+(\inot)\, \setminus\, \inot)$ of $M$, we obtain a manifold with boundary that serves as the conformal completion of $(\hat{M},\, \hat{g}_{ab})$ satisfying \emph{Definition 1}, with $\scrip = (\dot{J}^+(\inot)\, \setminus\, \inot)$. %Specifically, as noted in section \ref{s2.1}, the requirement that  $g_{ab}$ be $C^4$ at $\scrip$ ensures that the asymptotic Weyl curvature of the physical space-time has the Newman-Penrose `peeling property' there, and the leading order piece $K_{abcd}:= \lim_{\to \scrip}\,\Omega^{-1} C_{abcd}$ satisfies the Bianchi identities that provide us with balance equations for BMS momenta. However, as we remarked earlier, this assumption is made mainly to simplify the presentation and, as we will discuss in Section~\ref{s5}, it can be weakened. %Note furthermore that the peeling properties do \emph{not} restrict the behavior of the asymptotic Weyl curvature as one approaches $\inot$ along $\scrip$. In particular, the Newman-Penrose component $\Psi_1^\circ$ could well diverge in this limit, even when peeling holds on $\scrip$. 
Similarly, if we focus only on the conditions that $\Omega$ and $g_{ab}$ satisfy at $\inot$, we recover \emph{Definition 2} of section \ref{s2.3}, from which the standard ADM \cite{adm} and Geroch \cite{rg-jmp} 3+1 description follows by choosing space-like sub-manifolds $\Sigma$ of $M$ that are $C^{>1}$ at $\inot$ \cite{am3+1}. Thus, \emph{Definition 3} serves to unify the disparate notions of asymptotic flatness in null and spatial regimes. Note that, being $C^{>0}$ at $\inot$ the metric $g_{ab}$ is in particular continuous, irrespective of whether one  approaches $\inot$ along space-like directions or null. 

There is however a key difference relative to our discussion of null infinity in sections \ref{s2.1} and \ref{s2.2}. In that discussion we used the conformal freedom to make $\scrip$ divergence-free, i.e., to set $\nabla_a \no^a =0$, which via Einstein's equations, then implied $\nabla_a \no_b \hat= 0$ at $\scrip$. While this restriction was not essential, it simplified the subsequent discussion of null infinity considerably. However, this restriction on conformal frames is incompatible with the strategy of representing $\scrip$ as the future null cone of $\inot$ since the generators of null cones have non-zero expansion (already in Minkowski space-time). But one can easily change the requirement on $\Omega$ to adapt to the null cone structure of $\scrip$ and ask that the vector field\, $n^a := \nabla^a \Omega$\, have \emph{constant, positive} divergence, on $\scrip$. If we set $\nabla_a n^a = 4\Phi$, Einstein's equations then imply $\nabla_a n_b = \Phi\, g_{ab}$ on $\scrip$. In view of the condition $\nabla_a\nabla_b \Omega = 2\, g_{ab}$ at $\inot$, in this section we will restrict ourselves to conformal factors $\Omega$ for which\, $\Phi \, \= \,2$\, on $\scrip$ in a neighborhood of $\inot$ \cite{aaam-prl}, so that 
\begin{equation}
    \nabla_a n_b := \nabla_a\, \nabla_b \Omega\,\, \hat=\,\, 2 g_{ab} \qquad \hbox{on $\scrip$ in this neighborhood}\, . 
\end{equation}
% �          
Given an AEFANSI space-time $(\hM, \hg_{ab})$, we can always choose such a conformal completion; there is no loss of generality.

How are these constant divergence conformal frames related to the divergence-free frames of sections \ref{s2.1} and \ref{s2.2} that are normally used in the literature on null infinity? Let us begin with a conformal completion of \emph{Definition 1} that endows $\scrip$ with a Bondi conformal frame, i.e. a conformal factor $\Omega$ for which $\nablao_a \no_b\, \=\,0$ and $\qo_{ab}$ is a unit 2-sphere metric. Let $\Omega$ be a conformal factor for which $\nabla_a n_b\, \= \,2 g_{ab}$ on $\scrip$ to the past of some cross-section (i.e. in a neighborhood of $\inot$). Then these two are conformal completions are related by $\Omega = \omegao \Omegao$ where $\omegao$ is $C^4$ everywhere except at $\inot$, vanishes at $\inot$ and is $C^{>0}$ there. The condition on divergence of $\no^a$ and $n^a$ is satisfied if and only if $\Lie_{\no} \omega^{-1} = -2$ on the portion of $\scrip$ under consideration. Thus, $\uo = - 1/(2\omegao)$ is an affine parameter of $\no^a$. Since cross-sections $\uo = {\rm const}$ have unit area radius in the Bondi-frame $(\qo_{ab}, \no^a)$, \,$\omegao$ has the geometrical interpretation as the (time changing) area-radius of these cross-sections with respect to $g_{ab} = \omegao^2 \go_{ab}$. An affine parameter $u$ of the null normal $n^a = \omegao^{-1} \no^a$ of the given constant divergence conformal frame is given by $e^{-2u} =-\uo$. (Here we have chosen the integration constants so that $u$ and $\uo$ are constant on the 2-spheres $\omegao = {\rm constant}$.) 

To summarize, the divergence-free and the diverging conformal frames are related by:
\be \label{omegao}  g_{ab} = \omegao^2 \go_{ab},\qquad {\rm with}\qquad \omegao^{-1}\, = \, (-2\uo)\, = \,2e^{-2 u} \ee
where $\uo, \, u$ are affine parameters of $\no^a, \, n^a$, respectively, which are constant on the $\omegao\, =\, {\rm const}$ cross-sections of $\scrip$. Both $n^a$ and $\omega$ vanish at $\inot$, and $\uo$ and $u$ tend to $-\infty$ there. This construction can be carried out starting with any divergence-free frame; we used a Bondi conformal frame because: (i) $\omegao$ has the geometrical interpretation as area radius w.r.t. $g_{ab}$ only if $\go_{ab}$ is a Bondi frames; and, (ii) the Newman Penrose framework is tied to these conformal frames.\vskip0.1cm

{To further probe the relation between the divergence-free and constant divergence conformal frames provided by $\go_{ab}$ and $g_{ab} = \omegao^2 \go_{ab}$, let us examine the relation between various fields in the two conformal frames. Consider a generic covector field $f_a$ on $\scrip$ that admits a non-vanishing direction dependent limit to $\inot$ (i.e., its components in any $C^{>0}$ chart admit non-vanishing direction dependent limits as one approaches $\inot$ along $\scrip$). Since $g_{ab}$ is $C^{>0}$ at $\inot$,  $f^a = g^{ab} f_b$ also admits a direction-dependent limit. %so is $f^a = g^{ab} f_b$. 
On the other hand, since $\go^{ab} = \omegao^{2} g^{ab}$, setting $\mathring{f}^a = \go^{ab} f_a$ we find that $\mathring{f}^a$ must vanish as $\omegao^2 \,\sim \, 1/{\uo}^2$ in the approach to $\inot$. Now, $\mo_a,\, \bar{\mo}_a,\,\lo_a := \D_a \uo$\, is a null co-triad on $\scrip$ that is adapted to $\go_{ab}$ in the sense that the only non-vanishing inner product between them is $\go^{ab} \mo_a \bar{\mo}_b = 1$. A corresponding co-triad adapted to $g_{ab}$ that admits non-zero limits to $\inot$ is  $ m_a = \omegao \mo_a,\, \bar{m}_a = \omegao \bar{\mo}_a,\,\, \omega_a:= D_a \omegao$. Since $f^a$ admits a non-zero direction dependent limit to $\inot$, so do its components $f^a m_a,\, f^a \bar{m}_a$ and $f^a \omega_a$. On the other hand, since $\omegao = -{1}/(2{\uo})$ it follows that the three components of $\fo^a$ in the triad adapted to $\go_{ab}$ have different behaviors: generically $\mathring{f}^a \lo_a$ admits a non-zero limit (that equals half the limit of $f^a D_a \omegao$), while $\mathring{f}^a \mo_a$ vanishes as $1/\uo$. The relation between more general fields $\mathring{f}_{a\ldots b}$ and $f_{a\ldots b}$, adapted to the divergence-free and constant divergence frames, can be readily derived from that for 1-forms. We will see that, in \emph{AM}  space-times, this relation naturally leads to the $1/\uo^n$-type falls-off conditions  on ${}^\star\Ko_{ab}$ and the news tensor $\No_{ab}$ that are normally assumed at $\scrip$, in addition to \emph{Definition 1}. }

\subsection{Asymptotically Minkowski space-times}
\label{s3.2}
\begin{figure}
    \centering
    \includegraphics{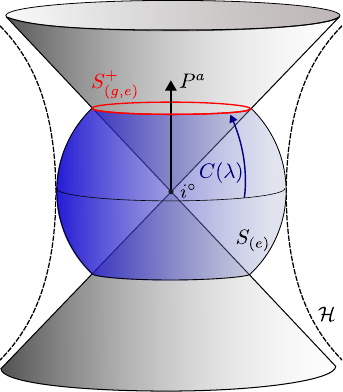}
    \caption{\emph{Tangent space $T_{\inot}$}: The (gray) cones are the future and past light cones of $\inot$ in $T_{\inot}$, and (the blue) $S_{(e)}$ is (a portion of) the 3-sphere spanned by vectors that are unit w.r.t. the Euclidean 4-metric $e_{ab}$, constructed by using the preferred timelike direction along the ADM 4-momentum $P^a$. $S^+_{(g,e)}$ (in red) is the 2-sphere intersection of $S_{(e)}$ with the future null cone.  $C(\lambda)$ describes a curve in the euclidean 3-sphere $S_{(e)}$ that approaches the 2-sphere $S^+_{(g,e)}$ as $\lambda\to 1$. The dashed line $\mathcal{H}$ is the hyperboloid of unit spacelike directions with respect to the Lorentzian metric $g_{ab}$. }
\label{fig:tangent-io}
\end{figure}
To arrive at the notion of \emph{AM} space-times, we need to impose compatibility conditions between limits of appropriate fields as one approaches $\inot$ along space-like directions and along $\scrip$. But there is a technical difficulty that requires us to make a small detour. Recall from section \ref{s2.3} and \ref{s2.4} that, in the discussion of spatial infinity, fields $T_{a\ldots b}{}^{c\ldots d}$ of physical interest admit regular direction dependent limits $\mathbf{T}_{a\dots b}{}^{c\ldots d} (\eta)$ at $\inot$, where $\eta^a$ is \emph{unit} vector in the tangent space of $T_{\inot}$ of $\inot$, tangential to the space-like curve used to take the limit. On the other hand, along $\scrip$ one approaches $\inot$ following the integral curves of its null normal $n^a$ which \emph{vanishes} at $\inot$. Therefore, $n^a$ cannot serve as the label of the direction of approach. One might imagine simply rescaling $n^a$ appropriately and using instead $n^{\prime\,a} = f(u) n^a$ that has a non-zero (direction-dependent) limit to $\inot$. However, even this $n^{\prime\,a}$ cannot be the limit of a 1-parameter family of `boosted' $\eta^a$ since $g_{ab} \eta^a \eta^b =1$ for the entire family, while $g_{ab}n^{\prime\, a}  n^{\prime\, b} =0$ (and $g_{ab}$ is continuous at $\inot$). Consequently, we cannot use it to specify the sense in which the limit $\mathbf{T}_{a\dots b}{}^{c\ldots d} (\eta)$ of a field ${T}_{a\dots b}{}^{c\ldots d}$ on $M$ is to be compatible with the limit taken along $\scrip$. 

A possible strategy to take the limit is to first note that the limiting fields $\bf{T}_{a\ldots b}{}^{c\ldots d}(\eta)$ can be naturally thought of as living on the unit time-like hyperboloid $\H$ in the tangent space of $\inot$ and $\H$ becomes \emph{asymptotically} tangential to the future null cone at $\inot$ in the distant future. Therefore, one can envisage making another conformal completion, now of $\H$, and identifying the future-directed null directions in $T_\inot$ with points on the future boundary of this completion  (see, e.g. \cite{kpis}). This would correspond to the intuitive idea of arriving the null directions by performing infinite boosts on space-like ones. However, the additional conformal completion of $\H$ makes the procedure rather cumbersome and one also has to check that the final results are independent of the choice of that completion. We will avoid these complications by using the following alternate strategy.

Recall, first, that the ADM 4-momentum $\admP$ is a time-like vector in the tangent space $T_\inot$ of $\inot$. Denote by $\tau_a$ the unit covector, $\tau_a = \admP/m_{\rm ADM}$, along $\admP$ and introduce an invariantly defined  positive definite---or, Euclidean---metric $e_{ab}$ on $T_{\inot}$:\, $e_{ab} := g_{ab} + 2 \tau_a \tau_b$. Consider the \emph{unit} 3-sphere ${S}_{(e)}$ defined by $e_{ab}$ in $T_\inot$, and denote by  $S^\pm_{(g,e)}$ the 2-sphere intersections of the null cone of $g_{ab}$ with ${S}_{(e)}$. Then each null generator of $\scri^\pm$ defines points $n^\pm$ on $S^\pm_{(g,e)}$, while each space-like tangent vector at $\inot$ defines a point in the open region  
of ${S}_{(e)}$ bounded by $S^\pm_{(g,e)}$. We will denote by $c(\lambda)$ smooth curves on the 3-sphere ${S}_{(e)}$ such that, for $\lambda \in (-1,\, 1)$ the point $c(\lambda)$ defines a space-like direction $\eta$, and for $\lambda =\pm 1$ null directions $n^\pm$\, (where space-like and null refer to $g_{ab}$). See Fig.~\ref{fig:tangent-io} for an illustration. Thus, each curve $c(\lambda)$ defines a 1-parameter family of space-like directions $\eta(\lambda)$ approaching $\inot$ that tend, in the limit $\lambda \to 1$, to a null direction $n^+$ approaching to $\inot$ along $\scrip$. Therefore, each tensor field ${T}_{a\ldots b}{}^{c\ldots d}$ on $M$ that admits a regular direction dependent limit to $\inot$ now provides us with a 1-parameter family of tensors $\mathbf{T}_{a\ldots b}{}^{c\ldots d} (\lambda)$ for $\lambda \in (-1,1)$. Similarly, the limit to $\inot$ of ${T}_{a\ldots b}{}^{c\ldots d}$ along $\scrip$ provides us with a tensor $\mathbf{T}_{a\ldots b}{}^{c\ldots d}\mid_{\lambda =1}$ at $\inot$ (which, a priori, could diverge). The idea is to tie the limits along space-like directions to those along null directions  by imposing a continuity requirement as $\lambda \to 1$ for appropriately chosen fields. In what follows this auxiliary structure stemming from the Euclidean metric $e_{ab}$ is used \emph{only} to specify the  topology in the space of directions in $T_{\inot}$ in a concrete fashion to make the notion of ``continuous families of space-like directions approaching null directions" explicit. 
\footnote{ Alternatively, following the procedure used in \cite{rmw-book} to explain the notion of $C^{>n}$ structures in the space-like approach to $\inot$, one can first choose $C^{>1}$ coordinates $x^{\mathfrak{a}}$ in a neighborhood of $\inot$ and, regarding them as `Cartesian' coordinates in a 4-d Euclidean space, introduce `angular' coordinates $(\theta,\varphi, \chi)$. Then $(\theta,\varphi, \chi)$ can then be used, in place of our 3-sphere $S(e)$, to specify how a family of space-like directions approaches a null direction. We chose a more streamlined procedure: $S(e)$ is constructed using the Euclidean metric $e_{ab}$ at $\inot$ which is defined \emph{invariantly} in any given  space-time, even though the ADM 4-momentum varies from one space-time to another.}
With this structure at hand, let return to the problem of `gluing' structures at spatial and null infinity appropriately. Since the metric $g_{ab}$ is $C^{>0}$ at $\inot$, its limit along the space-like directions is the same as that along null directions. Furthermore, in AEFANSI space-times, the connection $\nabla$---or, alternatively, the Christoffel symbols it defines in $C^{>1}$ charts---is required to admit regular direction limit along space-like directions. We will now extend that condition to include null directions by introducing a
\begin{quote}
\emph{Strengthening of the $C^{>0}$ requirements on the metric:}\\
We will require that as one approaches $\inot$ along $\scrip$, the first derivatives of the components of the metric in any $C^{>1}$ chart in a neighborhood of $\inot$ should admit direction dependent limits that are smooth in angles $(\theta,\varphi)$. This provides fields that are smooth on the 2-sphere $S^+_{(g,e)}$. Since derivatives of the metric components already admit regular direction dependent limits along space-like directions in all AEFANSI space-times, we also have smooth fields in the open region of the 3-sphere $S_{e}$ corresponding to space-like directions in $T_{\inot}$. Our `gluing'  requirement is that these fields be continuous at $S^+_{(g,e)}$ along any smooth curve $C(\lambda)$ on $S_{e}$. 
\end{quote}
%
%\footnote{To impose this condition, we have to strengthen the $C^{>1}$ charts on the manifold in the obvious manner: the coordinate derivatives of the transition maps now have direction dependent limits along spacelike \emph{and} null directions, and are continuous on $S^+_{(g,e)}$.}
%

Thus, in any $C^{>1}$ chart, not only do the Christoffel symbols---and hence the connection $\nabla$---admit regular direction limit along space-like directions, but they also admit direction-dependent limits as one approaches $\inot$ along $\scrip$ that are smooth in angles $(\theta,\varphi)$, and continuous at the 2-sphere $S_{(g,e)}$. (To impose this condition, we have to strengthen the $C^{>1}$ charts on the manifold in the obvious manner: the coordinate derivatives of the transition maps now have direction dependent limits along spacelike \emph{and} null directions, and are continuous on $S^+_{(g,e)}$.) Note, however, that since we only require continuity of $\nabla$, this condition \emph{does not constrain the behavior of curvature tensors in the limit $\lambda\to 1$ along curves $c(\lambda)$}, i.e., as a family of space-like directions approaches a null direction. By contrast, the $C^{>0}$ requirement on $g_{ab}$ in the approach to $\inot$ along \emph{space-like} directions does imply that $\Omega^{\f{1}{2}}\, R_{abcd}$ admits regular direction dependent limits along \emph{those} directions. Thus, our strengthening of the $C^{>0}$ condition on the metric does \emph{not} require that ``$g_{ab}$ be $C^{>0}$ in space-like \emph{and} null directions". 

Next, let us consider the set $\mathcal{T}$ of tensor fields ${T}_{a\ldots b}{}^{c\ldots d}$ on $M$, that are:\,  (i) constructed from the conformal factor $\Omega$, the metric $g_{ab}$, and the connection $\nabla$;\, (ii) admit regular direction dependent limits, $\mathbf{T}_{a\ldots b}{}^{c\ldots d} (\eta)$, as one approaches $\inot$ along space-like directions; and,\, (iii) are $C^1$ in a neighborhood of $\scrip$. Simplest examples of elements of $\mathcal{T}$ are $g_{ab}$ and $g^{ab}\,(\nabla_a \Omega^{\f{1}{2}})\, (\nabla_b \Omega^{\f{1}{2}})$.
\footnote{In addition, elements of $\mathcal{T}$ also include integrals over, e.g., 2-spheres $\hat{S}$ in a continuous family that converges to smooth cross-sections in the limit to $\scrip$, and shrink to a single point in the approach to $\inot$ (the 2-sphere integrals being performed using the volume element induced by $g_{ab}$).  A simple example is provided by functions $\oint_{\hat{S}} \epsilon_{ab} \equiv \oint_{\hat{S}}  \rmd^2{V}$\, where $\epsilon_{ab}$ is the volume 2-form on the 2-spheres $\hat{S}$ induced by $g_{ab}$.} 
To bring out the non-triviality of the conditions that define $\mathcal{T}$, let us list a few simple candidates that fail to belong to $\mathcal{T}$. Note first that $\Omega^{-1}\, C_{abcd}\,\nabla^b \Omega \nabla^d\Omega$ is $C^1$ at $\scrip$, it does not belong to $\mathcal{T}$ because it does not satisfy condition (ii). Similarly, although $\nabla_a \Omega^{\f{1}{2}}$\, and\, $\Omega^{-\f{1}{2}} C_{abcd} \nabla^b \Omega \nabla^d\Omega$ admit regular direction dependent limits to $\inot$, they  do not belong $\mathcal{T}$ because they do not satisfy (iii).  Also, $\mathcal{T}$ does \emph{not} contain fields that refer to additional structures such as the $\omegao$ relating the divergence-free conformal frames to those that are well-behaved at $\inot$. Thus, the set $\mathcal{T}$ is quite small, whence the continuity condition we now introduce is quite weak. In essence, it is the ``weakest matching condition'' one needs to relate the two \Poincare groups $\bmsP$ and $\spiP$ in this paper and the associated charges and fluxes in the companion paper \cite{ak-J}.\\

The second `gluing condition' will be on tensor fields in $\mathcal{T}$. \vskip-1.4cm
{\begin{quote}\vskip-0.3cm {\emph{Continuity requirement}:\, Fields in $\mathcal{T}$ should admit a limit $\mathbf{T}_{a\ldots b}{}^{c\ldots d}(n^+)$ as one approaches $\inot$ along $\scrip$ and, furthermore, given any of our curves $c(\lambda)$ on the 3-sphere $\mathbf{S}_{(e)}$,\,\,the one parameter family of tensors $\mathbf{T}_{a\ldots b}{}^{c\ldots d} (\lambda)$ must satisfy:\,\,\, $\lim_{\lambda \to 1} \mathbf{T}_{a\ldots b}{}^{c\ldots d} (\lambda) =  \mathbf{T}_{a\ldots b}{}^{c\ldots d}\mid_{\lambda =1}$}. \end{quote}}
\vskip -0.1cm  \noindent Thus, the condition requires the elements of $\mathcal{T}$ to admit direction-dependent limits as one approaches $\inot$ along $\scrip$, and further asks that limits to $\inot$ along a 1-parameter family of space-lime directions should converge to the limit along the null direction to which the space-like directions converge.

The last asymptotic condition we will impose is the same as the one that led us to the reduction of the Spi group to the \Poincare group in section \ref{s2.4}: \,\, $ \lim_{\to \inot} \Omega^{\f{1}{2}}\, {}^\star C_{abcd}\, \eta^b\eta^d =0$ as one approaches $\inot$ along space-like directions, or, equivalently, $\beta_{ac} := {}^\star C_{abcd}\, \eta^b\eta^d$ should admit a regular direction dependent limit as one approaches $\inot$ in space-like directions. We will rewrite this requirement using tensor fields that belong to $\mathcal{T}$:
\begin{quote}{\emph{Fall-off of the magnetic part of the Weyl tensor:}\, The field\, ${}^\star K_{ac} := \Omega^{-1}\,{}^{\star} C_{abc}{}^d \, \nabla^b\Omega\, \nabla_d \Omega$ should admit a regular direction dependent limit to $\inot$ along space-like directions. (Using the fact that $\eta^a = \lim_{\to \inot}\, \nabla_a \Omega^{\f{1}{2}}$, it follows that the limit is $4\,\bfbeta_{ac}(\eta)$.)}\end{quote}
\vskip0.1cm
Together, the three gluing conditions, discussed above, lead us to the notion of Asymptotically Minkowski space-times:\medskip

\textbf{Definition 4:}\, An AEFANSI space-time $(\hat{M}, \hat{g}_{ab})$ will be said to be \emph{Asymptotically Minkowskian} (\emph{AM}) if there exists a completion in which the above three requirements are met: the metric $g_{ab}$ satisfies the strengthened $C^{>0}$ condition; (ii) fields in $\mathcal{T}$ satisfy the continuity requirement; and, (iii) the magnetic part of the Weyl tensor satisfies the stronger fall-off as one approaches $\inot$ along space-like directions.\vskip0.1cm \goodbreak

{\it Remarks:} 

(1) It is instructive to examine the continuity requirement using various fields in $\mathcal{T}$. Consider first $ f:= g^{ab}\, \nabla_a\Omega^{\f{1}{2}}\,\nabla_b \Omega^{\f{1}{2}}$. As one approaches $\inot$ along \emph{any} $C^{>1}$ space-like curve, $f \to 1$. As we approach $\scrip$, we have $f\, \=\, \lim_{\scrip}\, (1/4\Omega)\, g^{ab}\, \nabla_a \Omega\, \nabla_b\Omega$, which is finite since the numerator goes to zero as $\Omega$. (Thus, although $ \nabla_a  \Omega^{\f{1}{2}}$ diverges in the limit at $\scrip$, its norm is well-defined.) We can use the L'H\^{o}pital rule to evaluate the limit as one approaches $\inot$ along $\scrip$. Using the equality $\nabla_a \nabla_b \Omega\,\, \=\,\, 2 g_{ab}$ at $\inot$, one finds that the limit is again $f \to 1$. Therefore, along any of our curves $c(\lambda)$, we have $\lim_{\lambda\to 1} f(\lambda) = f|_{\lambda=1}$. Thus, \ the continuity requirement is not an additional restriction for this simple element of $\mathcal{T}$; it holds already in AEFANSI  space-time.

(2) The situation is different for ${}^\star K_{ac} = \Omega^{-1}\,{}^{\star} C_{abc}{}^d \, \nabla^b\Omega\, \nabla_d \Omega$, which is also in $\mathcal{T}$ (thanks to the third requirement in \emph{Definition 4} on the magnetic part of the Weyl tensor). The continuity condition implies that this field must admit direction dependent limit to $\inot$ also along $\scrip$ (which is smooth in angles $\theta,\varphi$). This condition is not automatically met in AEFANSI space-times; it is a genuine restriction imposed in the passage to  \emph{AM} space-times.

(3) Let us return to the field $\Omega^{-\f{1}{2}}\, E_{bd} :=\Omega^{-\f{1}{2}} C_{abcd} \nabla^a\Omega \nabla^c\Omega$ we mentioned while introducing the set $\mathcal{T}$ of fields that are subject to the continuity requirement. We can rewrite it as $\Omega^{\f{1}{2}}\, K_{abcd}\, n^a n^c$, whence it is continuous at $\scrip$ and vanishes identically there. In particular, then, its limit to $\inot$ along $\scrip$ also vanishes. Along space-like approach to $\inot$, the field admits a regular direction dependent limit, $\f{1}{4}\, \mathbf{E}_{bd}$, which is a smooth field on the hyperboloid $\mathcal{H}$ in the tangent space $T_{\inot}$ of $\inot$, satisfying a deterministic evolution equation. However, a detailed examination of its asymptotic behavior shows the limit of $\mathbf{E}_{bd}$ diverges along a general 1-parameter family of  space-like directions that continuously converges to a null direction! This is a concrete example of a field which has good behavior as we approach $\inot$ along both in null and space-like directions \emph{separately,} but the two sets of limits do not match harmoniously: there is an infinite discontinuity in the limit in which space-like directions approach null directions. Note that there is no inconsistency with our continuity requirement because  $\Omega^{-\f{1}{2}}\,{E}_{bd}$ is not in $\mathcal{T}$: It is $C^0$ at $\scrip$ but not $C^{1}$. This illustrates the necessity of $C^{1}$ regularity at $\scrip$; we could not have broadened $\mathcal{T}$ by asking only continuity condition near $\scrip$! It also shows that the gluing procedure is rather delicate because there can be unforeseen, subtle constraints. 

(4) Since the main results of this paper refer to the relation between $\scrip$ and $\inot$, we left out a discussion of the approach to $\inot$ along $\scrim$. But it is clear from the above discussion (and the discussion that will follow), that all our considerations can be extended in a straightforward manner to $\scrim$. Note, however, that the relevant Weyl curvature on $\scripm$ is $K_{abcd} \equiv \Omega^{-1}\, C_{abcd}$,\, which is not in our collection $\mathcal{T}$ because it does not admit direction dependent limits to $\inot$ in space-like directions. Therefore, our assumptions do not imply that the past limit of $K_{abcd}$ along $\scrip$ and future limit along $\scrim$ exist and are continuous, as is sometimes assumed in the literature. Indeed, in generic \emph{AM} space-times the limits diverge.

\section{The Poincar\'e reduction in \emph{AM} space-times}
\label{s4} 

This section is divided into 3 parts. In the first, we show that in any \emph{AM} space, ${}^\star{\Ko}_{ab}$ and $\No_{ab}$ are guaranteed to have the fall-off properties that are normally imposed at $\scrip$ in addition to \emph{Definition 1}. In the second and third parts we show that the asymptotic symmetry group of \emph{AM} space-times is a \Poincare group $\inotP$ that coincides with the preferred subgroup $\bmsP$ of $\BMS$ discussed in section \ref{s2.2}, and with the preferred subgroup $\spiP$ of $\Spi$ of section \ref{s2.4}. In the companion paper \cite{ak-J} we will show that in these space-times the physically expected relation between angular momentum at $\scrip$ and $\inot$ holds for generators of $\inotP$ without further assumptions, e.g., on the limits of fields that enter their integrands.

\subsection{Curvature fall-off in the approach to $\inot$ along $\scrip$}
\label{s4.1}

Conditions (ii) and (iii) in the \emph{Definition 4} of \emph{AM} space-times  ensure that ${}^\star K_{ac} = \Omega^{-1}\,{}^\star{C}_{abc}{}^{d}\, n^b\,n_d$ admits direction dependent limits\, ${}^\star K_{ac} (n^{+})$\, as one approaches $\inot$ along $\scrip$. Let us translate this property to a divergence-free conformal frame $\go_{ab} = \omegao^{-2} g_{ab}$.\, Because $\Omega = \omegao\,\Omegao$,\, $K_{abc}{}^d = \Omega^{-1} C_{abc}{}^d$, and $C_{abc}{}^d$ is conformally invariant, we have ${}^\star{\Ko}_{ac} = \omegao\, {}^\star{K}_{ac}$. Since $\omegao = -1/2\uo$, this implies that ${}^\star{\Ko}_{ac}$ falls-off (at least as fast) as $1/\uo$,\, in the limit $\uo \to -\infty$ in divergence-free conformal frames. As we saw in section \ref{s2.2}, this is precisely the condition that led to the reduction of the BMS group $\BMS$ to the \Poincare group $\bmsP$ in \cite{aa-rad}. Thus, the continuity requirement together with the assumption on fall-off of the magnetic part of the Weyl curvature in \emph{space-like directions} implies that the necessary condition for the \Poincare reduction  \emph{at null infinity by itself} is automatically satisfied.

Let us translate these considerations to the Newman-Penrose notation. As we discussed at the end of section \ref{s3.1}, since $\omegao^{-1} {}^\star{\Ko}_{ac}$ admits a limit, so does  $\omegao^{-5} {}^\star{\Ko}^{ac}$. Since this ${}^\star{\Ko}^{ac}$ is tangential to $\scrip$ (i.e., ${}^\star{\Ko}^{ac}\no_a\, \=0$), it suffices to restrict oneself to its components in the triad $\lo_a, \mo_a, \bar{\mo}_a$ and examine their fall-offs as one approaches $\inot$ along $\scrip$. Using the relation between these triad co-vectors and $\omega_a = D_a\omegao,\, m_a,\, \bar{m}_a$ that have well-defined, non-vanishing limits to $\inot$, one concludes that $\omegao^{-1} {}^\star{\Ko}^{ac}\lo_a\lo_c \equiv \omegao^{-1}{\rm Im}\Psi_2^\circ$, \,  $o^{-2} {}^\star{\Ko}^{ac}\lo_a \bar{\mo}_c \equiv \omegao^{-2}\Psi_3^\circ$,\, and\, $\omegao^{-3} {}^\star{\Ko}^{ac}\bar{\mo}_a \bar{\mo}_c \equiv \omegao^{-3}\Psi_4^\circ$\, also admit well-defined, direction dependent limits as one approaches $\inot$ along $\scrip$. Hence, in the divergence-free conformal frames we have ${\rm Im} \Psi_2^\circ \sim O(1/\uo)$,\, $\Psi_3^\circ \sim O(1/\uo^2)$,\, and  $\Psi_4^\circ \sim O(1/\uo^3)$. The fall-off of ${\rm Im} \Psi_2^\circ$ is precisely the one used in \cite{etnrp} to arrive at a \Poincare reduction of the BMS group.\\

{\emph{Remark:} Interestingly, the relation between the Bondi news and $\Psi_3^\circ,\, \Psi_4^\circ$ immediately implies that $\No =\No_{ac}\mo^a \mo^c$ falls off as $1/\uo^2$. As we saw in section \ref{s2.3}, the fall-off condition of $\No$ is precisely the one that are normally imposed by hand, invoking physical considerations such as finiteness of the flux of energy momentum and angular momentum across $\scrip$.  The fact that this fall-off is automatically satisfied in \emph{AM} space-times provides non-trivial evidence that, although the strengthening of the AEFANSI conditions in \emph{Definition 3} was motivated primarily by geometrical considerations, it automatically captures correct physics.  Finally, we would like to clarify a conceptual issue. In the community working on asymptotics, it is often implicitly assumed that \emph{Definition 1}, together with the requirement that the limit of $B_{ab}$ should vanish as one approaches $\inot$ in \emph{space-like directions} automatically ensures these fall-off conditions on ${\rm Im} \Psi_2^\circ$,\, $\Psi_3^\circ$ and  $\Psi_4^\circ$ as $u\to -\infty$ on $\scrip$. This is not the case. One needs, in addition, the continuity condition (ii) in the \emph{Definition 4} of Asymptotically Minkowski space-times.
\\

To summarize, the necessary conditions that are normally imposed to arrive at \Poincare \\reductions of the BMS and Spi group are satisfied in \emph{AM} space-times. Therefore, we can use the procedure outlined in section \ref{s2.2} to reduce the BMS group to its \Poincare subgroup $\bmsP$, and that outlined in \ref{s2.4} to reduce the Spi group to its \Poincare subgroup $\spiP$. The question is: do we have a single asymptotic symmetry group $\inotP$ that coincides with $\bmsP$ at $\scrip$ and with $\spiP$ at $\inot$? Put differently, if we have a vector field $\xi^a$ in the physical space-time that belongs to the Lie algebra of $\spiP$, does it automatically belong to the Lie-algebra of $\bmsP$? We will now show that the answer is in the affirmative. For convenience of the reader, as in section \ref{s2.2}, we will provide two arguments. The first, based on connections $\{\Do\}$ on $\scrip$, is intrinsic and more direct in the sense that it does not involve additional structures such as null tetrads. The second is based on the asymptotically shear-free cross-sections used in \cite{etnrp}.%  

\subsection{Poincar\'e reduction using connections}
\label{s4.2}
Let us begin by recalling the reduction of the Spi group $\Spi$ to its \Poincare subgroup $\spiP$. Thanks to our condition on the magnetic part of the Weyl tensor in \emph{Definition 3}, one can select a \emph{preferred} subfamily of conformal completions $g_{ab}$, where any two metrics $g_{ab}$ and $g^\prime_{ab}$ are related by $g^\prime_{ab} = \omega^2 g_{ab}$ where $\omega\, \=\, 1$ at $\inot$ and $\nabla_a \omega$ admits a continuous limit to $\inot$. Recall that these preferred conformal completions and the corresponding metrics are referred to as \emph{admissible}.  Two admissible metrics are said to be equivalent if and only if\, $\nabla_a \omega\, \= \,0$ at $\inot$, and each equivalence class $\{g_{ab}\}$ is called a `ripple' at $\inot$.
%Each admissible completion provides us with a `ripple' $\{g_{ab}\}$ at $\inot$, where $g_{ab}$ and $g^\prime_{ab}$ belong to the same ripple if and only if\, $\nabla_a \omega\, \= \,0$ at $\inot$.
Thus, the difference between any two ripples in the preferred family is completely characterized by the covector $\nabla_a \omega$ at $\inot$.

Next, recall from Remark 3 at the end of Section \ref{s2.4} that each ripple $\{g_{ab}\}$ is in turn characterized by the equivalence  class $\{ \nabla_{\inot}  \}$ of asymptotic connections, where $\nabla_{\inot}$ is the derivative operator of an admissible metric in that ripple. $\spiP$ is the subgroup of $\Spi$ that  preserves the preferred family of ripples $\{g_{ab}\}$ or, equivalently, maps any $\{ \nabla_{\inot}  \}$ in the preferred family to another $\{ \nabla_{\inot}^\prime  \}$ in the preferred family via (\ref{conformal3}).

Now, thanks to our strengthening of the $C^{>0}$ requirement, in an \emph{AM} space-time, the connection $\nabla$ compatible with any admissible metric $g_{ab}$ is well-defined also on $\scrip$. Therefore, if we make a conformal transformation $\go_{ab} = \omegao^{-2} g_{ab}$ to pass to a divergence-free frame on $\scrip$, the connection $\nabla$ of an admissible $g_{ab}$ provides an equivalence class of intrinsic connections $\{ \Do \}$ on $\scrip$ such that the ${}^\star\Ko_{ab}$ defined by $\{\Do \}$ vanishes in the distant past (i.e., as $\uo \to -\infty$). Therefore, past limit of $\{ \Do\}$ provides us with a vacuum $\{\vac\}$ on $\scrip$, which transforms via (\ref{conformal2}) under further conformal rescalings $\go^\prime_{ab} = \muo^2 \go_{ab}$ to another divergence-free frame. Thus, the family of preferred $\{ \nabla_{\inot}  \}$ obtained from admissible completions --and therefore preserved under the action of $\spiP$-- induces a unique family of vacua $\{\vac\}$ on $\scrip$ in divergence-free completions. ($\nabla$ transforms via (\ref{conformal3}) as we move from one admissible metric to another, while $\{\vac\}$ transforms via (\ref{conformal2}) as we move from one the divergence-free conformal frame to another.) Now, since \emph{AM} space-times satisfy, in particular, \emph{Definitions 1} and \emph{2},\, every element of its symmetry group belongs to $\Spi$ as well as $\BMS$. If it belongs to $\spiP$, it preserves in addition the family  $\{ \nabla_{\inot} \}$ and hence by above reasoning, also the family $\{ \vac \}$. But the subgroup of $\BMS$ that preserves this family is precisely $\bmsP$. Thus, in \emph{AM} space-times, there is a single \Poincare group $\inotP$ whose action coincides with that of $\spiP$ at spatial infinity and with $\bmsP$ at null infinity. 

This completes our first argument, one based on connections. 

\subsection{Poincar\'e reduction using shear-free cross-sections}
\label{s4.3}

For the second argument, let us first recall (from Remark 1 at the end of Section \ref{s2.4}) that to reduce the Spi group $\Spi$ to $\spiP$  one restricts oneself to conformal completions such that ${\rm TF}\,\bfh_a{}^m\,\bfh_b{}^n{\bf{S}}_{mn} =0$, where TF stands for `trace-free part of',\,\, ${\bf{S}}_{mn} = \lim_{\to\inot} \Omega^{\f{1}{2}}\, S_{ab}$ (with $S_{ab}$, the Schouten tensor of $g_{ab}$),\, and\, $\bfh_{ab} = \lim_{\to\inot} h_{ab}$ (with $h_{ab}$ the intrinsic metric on the $\Omega={\rm const}$ 3-manifolds). These are the \emph{admissible} conformal completions that select the preferred set of ripples discussed above. The subgroup of $\Spi$ that preserves the family of admissible completions is precisely the \Poincare subgroup $\spiP$ of the Spi group $\Spi$.

Let us denote by $\ubS_{ab}$ the pull-back of $S_{ab}$ to the $\Omega={\rm const}$ 3-surfaces. Then, a completion is admissible if and only if ${\rm TF}\, \ubS_{ab}$ admits a regular direction dependent limit as one approaches $\inot$ along space-like directions. Our continuity condition in \emph{Definition 4} implies that, in any admissible completion, ${\rm TF}\, \ubS_{ab}$ also admits a direction dependent limit as we approach $\inot$ along $\scrip$. The question then is whether this condition also selects a \Poincare subgroup of the BMS group on $\scrip$, \emph{and} whether that group is precisely the $\bmsP$ of section \ref{s2.2} (selected by the family of asymptotically shear-free cross-sections). We will now show that the answers to these questions are in the affirmative.

Fix a physical space-time $(\hat{M},\hat{g}_{ab})$ and a conformal completion $(M, g_{ab})$ thereof, satisfying \emph{Definition 4} and providing us with an admissible completion, for which $\nabla_a n_b\, \=\, 2 g_{ab}$ on $\scrip$ (in  a neighborhood of $\inot$). Let $\go_{ab} = \Omegao^2 \hg_{ab}$ be a conformal completion satisfying \emph{Definition 1} which endows $\scrip$ with a Bondi conformal frame, i.e.,  in which $\nablao_a \no_b\,\, \= \,\,0$ and $\qo_{ab}$ is a unit 2-sphere metric. The admissible metric $g_{ab}$ is conformally related to the metric $\go_{ab}$ via $g_{ab} = \omegao^2 \go_{ab}$, and our discussion in section \ref{s3.1} implies that $\omegao$ provides an affine parameter $\uo$ for $\no^a$ via $\uo = -1/(2\omegao)$. Since $g_{ab}$ and $\go_{ab}$ are conformally related, we can compute the relation between their Schouten tensors. It is straightforward to verify that their pull-backs ${\ubS}_{ab}$ and ${\ubSo}_{ab}$ to $\scrip$ are related by (see e.g., \cite{kpis})
\be \label{S} {\rm {TF}}\,{\ubS}_{ab}\,\, \= \,\,{\rm {TF}}\,{\ubSo}_{ab}\, +\, 2\,{\rm TF}\,\, \omegao \Do_a \Do_b \omegao^{-1} \,\, \= \,\, {\rm {TF}}\,{\ubSo}_{ab}\, -\,  {2}\,{\uo}^{-1}\,{\rm{TF}}\,(\Do_a\,\ello_b )\, ,\ee
where $\lo_a = -\Do_a \uo $ is the covariant normal to the $\uo={\rm const}$ cross-sections of $\scrip$, with normalization $\no^a \lo_a = -1$. Now, because $g_{ab}$ provides an admissible conformal completion and satisfies the continuity requirement of \emph{Definition 4}, the left side of (\ref{S}) admits a direction dependent limit as we approach $\inot$ along $\scrip$. By Eq. (\ref{S}), then, the right side must have the same property. Since the right side is trace-free and transverse to $\no^a$, only its projection by $\mo^a\mo^b$ is non-trivial. We can now repeat our discussion on the behavior of ${}^\star\Ko_{ab}$ in the distant past to conclude that this projection must falloff as $1/\uo^2$ as $\uo \to \infty$. Thus, the fact that the left-hand side of (\ref{S}) admits direction dependent limits at $\inot$ implies that the right side must fall off as $1/\uo^2$ in a Bondi conformal frame.

Let us analyze the implications of this fact for the two terms on the right side. Since $\qo_{ab}$ is a Bondi conformal frame, ${\rm TF}\,\ubSo_{ab}$ is the Bondi news $\No_{ab}$,  while ${\rm{TF}}\, (\Do_a\,\ello_b )$ is the shear $\sigmao_{ab}$ of the $\uo ={\rm const}$ cross-sections in the Bondi frame $\go_{ab}$. Now, as shown above, in \emph{AM} space-times, $\No_{ab}\mo^a \mo^b$ is guaranteed to fall-off as $1/\uo^2$. Therefore, the second term should also fall-off as $1/\uo^2$. Since the term has only $1/\uo$ as coefficient of $\sigmao_{ab}$ we conclude that the shear $\sigmao = \sigmao_{ab} \mo^a\mo^b$ of the 1-parameter family of cross-sections $\uo = {\rm const}$ must fall off as $1/\uo$. To summarize, the fact that the left side of (\ref{S}) admits direction dependent limits as we approach $\inot$ along $\scrip$ implies that the foliation of $\scrip$ provided by $\uo = {\rm const}$ cross-sections is asymptotically shear-free. Here $\uo$ is an affine parameter of the $\no^a$ in the given Bondi frame.

What happens if we keep $g_{ab}$ fixed but change the Bondi conformal frame $\go_{ab} \to \go^\prime_{ab}$? Since the above reasoning only assumed that we are in a Bondi conformal frame, the foliation $\uo^\prime = {\rm const}$ is again asymptotically shear-free in the distant past. Since each Bondi frame $\go^{\prime}_{ab}$ is related to a fiducial $\go_{ab}$ by a boost, there is a 3-parameter set of Bondi frames. Thus, we obtain 3-parameter worth of foliations that become asymptotically shear-free in the distant past. (Within each foliation, cross-sections are related by a BMS time translation). The transformation property of shear under conformal rescaling implies that these cross-sections become asymptotically shear-free in every Bondi frame. Finally, let us change the admissible metric from $g_{ab}$ to ${\underline{g}}_{ab}$, related to the initial metric $\go_{ab}$ in the Bondi frame via ${\underline{g}}_{ab} = \underline{\omegao}^2 \go_{ab}$. Then we again obtain a 1-parameter family of cross-sections $\mathring{\underbar{u}} = {\rm const}$ that become asymptotically shear-free in the Bondi frame $(\qo_{ab}, \no^a)$, where $\mathring{\underbar{u}}$ is again an affine parameter of $\no^a$. Therefore, the two families $\uo = {\rm const}$ and $\mathring{\underbar{u}} = {\rm const}$ are related by a supertranslation. Now the transformation property of shear under supertranslation implies that the asymptotic shear can vanish for the two families if and only if the supertranslation is a BMS translation. Thus, the family $\mathring{\underbar{u}} = {\rm const}$ of asymptotically shear-free cross-sections obtained from ${\underline{g}}_{ab}$ is already contained in the set of cross-sections we obtained by varying $\go_{ab}$, keeping $g_{ab}$ fixed. 

To summarize, if a space-time is \emph{AM} then $\scrip$ is naturally endowed with a 4-parameter set of cross-sections that are asymptotically shear-free in any Bondi conformal frame. Therefore, this set must be left invariant by each element of $\spiP$ that preserves the set of admissible completions. Recall that this is precisely the condition that arises in the Newman-Penrose approach \cite{etnrp} to the selection of a canonical \Poincare subgroup of $\B$.  Thus, the symmetry group of \emph{AM} space-times is a \Poincare group  $\inotP$ that coincides with $\spiP$ in the spatial regime and with $\bmsP$ in the null.\vskip0.1cm

\emph{Remark:} At first sight, arguments presented in sections \ref{s4.2} and \ref{s4.3} may appear to be rather different from one another because they use quite different structures/tools. But, at a conceptual level, in both cases one simply translates the implications of admissibility of conformal completions \emph{at $\inot$}, to structures in \emph{divergence-free conformal completions} used in the discussion $\scrip$ by itself. 

In section \ref{s4.2} the emphasis is on the fact that, thanks to the `gluing conditions', the space-time connection $\nabla$ of any admissible completion at $\inot$ induces a unique equivalence class $\{\Do\}$ of intrinsic connections  given  any given divergence-free conformal frame $(\qo_{ab}, \no^a)$ on $\scrip$. In \emph{AM} space-times, each $\{\Do\}$ tends to a classical vacuum $\{ \vac \}$ on $\scrip$. Thus, given an admissible conformal completion, we obtain a classical vacuum at $\scri$. As we saw in section \ref{s2.4}, the subgroup of the Spi group $\Spi$ that preserves the set of admissible asymptotic completions is the \Poincare group $\spiP$. \emph{Thanks to the gluing conditions,} this subgroup automatically preserves the corresponding set of classical vacua $\{ \vac \}$ %in divergence-free completions of a given \emph{AM} space-time $(\hM, \hg_{ab})$ 
(in addition to the universal structure at $\scrip$). Therefore, each element of this group is also an element of $\bmsP$.

In section \ref{s4.3} the emphasis is on shear-free cross-sections in Bondi conformal frames. Now, Remark 2 at the end of section \ref{s2.2} summarizes how the asymptotically shear-free cross-section used in the Newman-Penrose approach \cite{etnrp} can be extracted, in an invariant manner, starting from classical vacua $\{ \vac\}$. Therefore, the group that preserves the set of classical vacua $\{\vac\}$ that arise from all divergence-free completions of $(\hM, \hg_{ab})$ also preserves the set of asymptotically shear-free cross-sections on $\scrip$ \`{a} la Newman-Penrose. Thus, the \Poincare reduction of section \ref{s4.3} could have been obtained directly from that of section \ref{s4.2}. In particular, the restriction to the Bondi conformal frames of section \ref{s4.3} is not necessary; indeed, as we pointed out earlier, conformal transformation property of shear implies that the Newman-Penrose family of cross-sections is asymptotically shear free in any divergence-free conformal frame. Our discussion of section \ref{s4.3} provides a self-contained argument in the spirit of \cite{etnrp} that does not refer to the intrinsic connections $\{\Do\}$.% (approaching classical vacua $\{\vac\}$ in the limit to $\inot$ along $\scrip$).

\section{Discussion}
\label{s5}

We will begin with a summary and then make several remarks that put our results in a broader context and suggest directions for future work.

The notion of asymptotic flatness was first formulated using outgoing null surfaces in the physical space-times \cite{bondi,sachs}, and recast soon thereafter using the notion of (weak) asymptotic simplicity that uses null geodesics and conformal completions \cite{rp}. This reformulation made it possible to analyze asymptotic structures --especially properties of  gravitational waves-- using local differential geometry at the boundary $\scrip$. It was later realized that the entire analysis can be carried out, again using a conformal completion, but without invoking either outgoing null surfaces or geodesics \cite{rg}. In the spatial regime, analysis of the asymptotic properties of the gravitational field was first carried out using a 3+1 decomposition \cite{adm,rg-jmp} and was then extended to a 4-dimensional setting using several different but related approaches \cite{aarh,aa-ein,rbbs,aajr}. In section \ref{s2} we summarized the understanding that emerged from these frameworks. In each regime the asymptotic symmetry group is infinite dimensional, because of the enlargement of the translation subgroup of the \Poincare group to a supertranslation group. This enlargement introduces a supertranslation ambiguity in the definition of angular momentum in each regime. However, both groups, $\BMS$ and $\Spi$, can be reduced to canonical \Poincare subgroups $\bmsP$ and $\spiP$ by introducing additional assumptions on the behavior of the ``magnetic part" of Weyl curvature as one approaches $\inot$ along $\scrip$, and along space-like directions, respectively. The two reductions are carried out separately, each in its own regime without reference to the other. Therefore, to relate the notion of angular momentum $\scriJ$ at null infinity with $\inotJ$ at spatial infinity, one needs to bring together the two disparate descriptions. 

The notion of AEFANSI Space-times, as introduced and analyzed in \cite{aa-ein}, is well suited to unify results that were obtained separately in the two regimes. We used it as the point of departure. In section \ref{s3.1} we first discussed the relation between AEFANSI conformal completions that are well-behaved both at $\scrip$ and $\inot$, and the divergence-free conformal completions, normally used in the gravitational radiation theory, which are well-behaved only at $\scrip$. In section \ref{s3.2} we supplemented the boundary conditions satisfied by AEFANSI space-times with three additional requirements to arrive at the key notion of \emph{Asymptotically Minkowski} (\emph{AM}) space-times. The first two requirements serve to appropriately `glue'  $\scrip$ to $\inot$. In AEFANSI space-times, the metric $g_{ab}$ is assumed to be $C^{0}$, and its connection --or Christoffel symbols-- are assumed to admit regular direction dependent limits  (only) in the space-like approach to $\inot$. We extended this notion by requiring that the Christoffel symbols should admit direction dependent limits also as one approaches $\inot$ along $\scrip$. These limits are assumed to be smooth in angles $(\theta,\varphi)$ but only continuous in the `boost' direction. Therefore, the extension of the $C^{>0}$ differentiability is quite weak: it does not have direct implications on the behavior of curvature as one approaches $\inot$ along null directions. The second gluing condition imposes a continuity requirement on the class $\mathcal{T}$ of tensor fields $T_{a\ldots b}{}^{c\ldots d}$ that are constructed only from the conformal factor $\Omega$ and the metric $g_{ab}$, admit regular direction dependent limits $\bfT_{a\ldots b}{}^{c\ldots d}$ in space-like approach to $\inot$, and are $C^1$ in a neighborhood of $\scri$.% 
\footnote{As discussed in section \ref{s3.2}, $\mathcal{T}$ is a rather small set. For example, $C_{abcd}$ is $C^2$ in a neighborhood of $\scrip$, but does not belong to $\mathcal{T}$ because it does not admit regular direction-dependent limits to $\inot$ along space-like directions. Similarly, $E_{ac} := \Omega^{\f{1}{2}}\, C_{abcd}\, \nabla^b \Omega^{\f{1}{2}} \,\nabla^d \Omega^{\f{1}{2}}$ and $E := \Omega^{\f{1}{2}} S_{bd}\, \nabla^b \Omega^{\f{1}{2}}\, \nabla^d \Omega^{\f{1}{2}}$ admit regular direction dependent limits to $\inot$ along space-like directions but do not belong to $\mathcal{T}$ because they are not $C^1$ at $\scrip$.}
Consider a one parameter family of space-like directions $\eta(\lambda)$ in the tangent space $T_\inot$, that tend continuously to a null direction $n^+$ along $\scrip$, as $\lambda \to 1$. The continuity requirement asks that if $T_{a\ldots b}{}^{c\ldots d}$ is in $\mathcal{T}$, then  the limit of\, $T_{a\ldots b}{}^{c\ldots d}$ to $\inot$ along the generator of $\scrip$ defined by $n^+$ should agree with $\lim_{\lambda \to 1}\, \bfT_{a\ldots b}{}^{c\ldots d} (\lambda)$. Again, this is only a continuity requirement; there is no condition on derivative w.r.t. $\lambda$. {In the literature, it is often assumed that fields that enter the integrand of charge integrals admit limits as one approaches $\inot$ along $\scrip$. This is not necessarily the case in \emph{AM} space-times.} The third and the last condition is not new: It was already introduced to remove the supertranslation ambiguity in the spatial regime \cite{aarh}. It asks that the ``magnetic part" of the Weyl tensor fall-off faster by one power of ``$1/r$" as one approaches $\inot$ \emph{along space-like directions}, than what is guaranteed in AEFANSI space-times. As discussed towards the end of section \ref{s2.4}, this condition is natural from several different perspectives.

In section \ref{s4} we first showed that the fall-off conditions on Bondi news and the ``magnetic part" of the Weyl tensor -- or $\Psi^\circ_4,\, \Psi^\circ_3,\, {\rm Im} \Psi^\circ_2$-- that are normally imposed in divergence-free conformal frames the null regime, \emph{in addition} to \emph{Definition 1}, are automatically satisfied in \emph{AM} space-times. This property illustrates the fact that, although the new `gluing' conditions were motivated using geometrical considerations, they naturally incorporate the physically expected behavior. We then established the desired result that the asymptotic symmetry group of these space-times is a single \Poincare group $\inotP$ that coincides with $\bmsP$ and $\spiP$ in the respective regimes. (The subscript $\inot$ in $\inotP$ is a reminder that the continuity conditions that lead to this \Poincare group refer to $\inot$, rather than $i^\pm$). In the companion paper \cite{ak-J} we discuss the asymptotic properties that a vector field $\xi^a$ in $M$ must have to generate a symmetry that belongs to $\inotP$, and then show that charges at null and spatial infinity associated with these generators are related in the expected way. In general this agreement will not hold if, e.g., one were to define angular momentum at $\scrip$ using another \Poincare subgroup of the BMS group $\BMS$, obtained by supertranslating $\bmsP$.

The geometric framework presented in this article is suitable for description of compact gravitating systems that include bound binaries as well as those that scatter off one another. Its main merit is that it captures the intuitive idea of asymptotic flatness in both null and space-like regimes in a single stroke. As pointed out at the end of section \ref{s2.2}, the framework is sufficiently general to allow situations in which there Newman-Penrose scalar $\Psi_1^\circ$ (in a Bondi conformal frame) diverges as one approaches $\inot$ along $\scri$.  This incorporates binaries that scatter off one another.\\

\emph{Remarks:}\vskip0.1cm

(1) Isolated systems in classical general relativity are asymptotically flat both in null and spatial regimes. The main result of this paper is that, if the asymptotic conditions that have been used separately in the two regimes are glued in a natural fashion, one arrives at the notion of \emph{AM} space-times, and the asymptotic symmetry group of these space-times is just the \Poincare group. From the viewpoint of $\scrip$ alone, this \Poincare reduction of the BMS group is unnatural because one can also select a \Poincare subgroup $\mathfrak{p}^{\rm bms}_{i^+}$ of $\BMS$ by requiring that ${}^\star\Ko_{ab}$ go to zero as we approach $i^+$ along $\scrip$. Generically, this group would a different \Poincare subgroup of $\BMS$; it would be the same as $\bmsP$ if and only if the gravitational memory vanishes in the given space-time. 

However, for physically interesting isolated systems, such as  stars and black holes, we have asymptotic flatness only at $\inot$; not at $i^+$! Therefore, it is natural to exploit the available structure and reduce the asymptotic symmetry groups $\B$ \emph{and} $\Spi$ in the two regimes in a single stroke to arrive at $\inotP$. As mentioned above, $\inotP$ plays a crucial role in relating the physical charges defined separately in the null and spatial regimes. Note also that in the post-Newtonian (PN) calculations, one does have a single \Poincare group that plays important roles, e.g., in providing notions of energy-momentum and angular momentum and balance laws associated with them that heavily used in the PN analysis of compact binary coalescence \cite{lb-rev}. From a viewpoint that focuses on $\scri$ alone, the important role played by this group in the PN analysis seems puzzling. From the present perspective it is not: $\inotP$ can be thought of as the natural generalization to \emph{AM} space-times in full general relativity of the PN \Poincare group. Indeed, in \emph{AM} space-times, the asymptotic conditions normally used on the initial data on Cauchy surface $\Sigma$ automatically follow under the natural condition that $\Sigma$ extend to $\inot$ as a $C^{>1}$ sub-manifold. And in this case, the definitions of energy-momentum and angular momentum defined on $\Sigma$ using Cauchy data agree with those associated to $\inotP$.

Note however, that these considerations do \emph{not} imply that one can just forgo the full BMS group $\BMS$ and work exclusively with $\inotP$ at $\scrip$. The supermomenta that $\B$ provides are physically interesting observables, and so are the associated gravitational memory and soft charges. These observables have been  used in classical general relativity, e.g., to compare and improve candidate waveforms \cite{kkadl,kak}, and in quantum gravity 
to understand the origin of infrared issues at a non-perturbative level \cite{aa-asym,aa-bib}, and to explore the relation between the BMS group and soft theorems of perturbative quantum gravity \cite{as-book}. However, it is equally true that these successes do \emph{not} imply that ${\inotP}$ is physically unimportant or will not exist in physically interesting situations. Indeed, it is often present also in the analyses aimed at uncovering relation between fields on $\scrip$ and $\scrim$ that emphasize the role of the BMS group  (see, e.g., \cite{kpis,cgw,Capone:2022gme}). It's just that the fact that a preferred \Poincare subgroup exists when one joins $\scrip$ and $\scrim$ through $\inot$ is either not noticed or not emphasized. \medskip

(2) Let us expand on this point by discussing the relation between our approach and these investigations. The  primary goal of these investigations is to explore the relation between charges at $\scrip$ and $\scrim$, and spatial infinity features only as an intermediate step. By contrast our analysis focuses on the relation between null and spatial infinity and charges refer to the \Poincare group $\inotP$ that naturally emerges at the interface of the two regimes \cite{ak-J}. In our approach these  \Poincare charges on $\scripm$ are related simply because each of them equals the total charge at $\inot$ minus the flux between $\inot$ and any given cross-sections of $\scripm$. 

In \cite{Prabhu:2019fsp,kpis}, spatial infinity is treated using the AEFANSI framework of \cite{aa-ein}, as in our approach. However,  in addition, the unit hyperboloid $\H$ in the tangent space at $\inot$ is conformally compactified using a second conformal factor. The future and past boundaries of this compactified hyperboloid corresponds to the null directions defined by the future and the past light cones at $\inot$. The second conformal factor features in the conditions that relate the limits of various fields as one approaches $\inot$ along $\scrip$, and the result of first taking limits to $\inot$ along space-like directions and then taking a second, infinite boost limit. Conditions imposed  on the Schouten tensor in \cite{kpis} are the same as in our treatment, whence there is a \Poincare subgroup $\spiP$ of $\Spi$ as well as a \Poincare subgroup $\bmsP$ of $\BMS$. However, the presence of these groups and their use is somewhat implicit.  In \cite{cgw,Capone:2022gme}, one works in the physical space-time rather than in its conformal  completion. Conditions at $\scri$ and $\inot$ are replaced by the assumption that the physical metric admits a Bondi-Sachs expansion \cite{bondi,sachs} as one recedes from sources in null directions, and a Beig-Schmidt expansion \cite{rbbs} as one recedes in space-like directions. One then uses techniques from matched asymptotic expansions. The Beig-Schmidt expansion implies in particular that the asymptotic symmetry group at spatial infinity is $\inotP$, but its presence is essentially ignored.

Our approach is more closely related to \cite{Prabhu:2019fsp,kpis} because both use conformal completions. However, there are also some important differences. First, our approach does not require a conformal completion of $\H$. Instead, the infinite boost limit is taken using the topology on the space of space-like and null directions which we  specify in a concrete fashion using an invariantly defined Euclidean metric in the tangent space $T_{\inot}$ of $\inot$.
%we introduce an invariantly defined Euclidean metric in the tangent space $T_{\inot}$ of $\inot$ and use it to endow the topology on the space of space-like and null directions at $\inot$ that is needed to discuss continuity of various fields in the `infinite boost limits'.  
Thanks to this strategy, we avoid the conceptual issues associated with uniqueness of the second compactification and technical complications in the subsequent equations. Second, our `gluing conditions' are weaker in the following sense. In \cite{Prabhu:2019fsp} the continuity condition is assumed also for some rescaled components of the Weyl tensor that appear in the integrands of charge integrals. These conditions then immediately imply the desired relations between supermomenta of spatial and null regimes. Our conditions in \emph{Definition 4} of \emph{AM} spacetimes do not imply these relations between the components of the Weyl tensor. In particular, in \emph{AM} space-times, the integrands can diverge in the infinite boost limit, but the charge integrals remain finite, as in \cite{kpis}. Finally, the asymptotic conditions used in \cite{Prabhu:2019fsp,kpis} are strong enough to relate charges associated with \emph{all} BMS symmetries on $\scri^\pm$, while in \emph{AM} space-times this result holds only for charges associated with symmetries in $\inotP$; there is no assurance that there is conservation of supermomentum between $\scri^\pm$. This difference and its relation to the issue of existence of solutions to full Einstein's equations---beyond perturbation theory---satisfying the boundary conditions necessary for supermomentum conservation are discussed in section 4 of \cite{ak-J}.  Our understanding of results to date \cite{kroon2} is that these conditions would be violated in generic scattering processes.
% Because of these differences the relation between charges at $\inot$ and at $\scri$ is not an immediate consequence of the `gluing conditions'. This point will become explicit in \cite{ak-J}.\medskip
 
(3) What is the status of existence of solutions satisfying the \emph{AM} boundary conditions? The post-Newtonian part of the standard analysis of compact binaries assumes that the system is stationary in the distant past \cite{lb-rev}. This implies that our condition on the magnetic part of the Weyl curvature is met. It should not be difficult to explicitly establish that our continuity condition that glues spatial and null regimes appropriately is also satisfied in this approximation. Similarly, numerical simulations of these systems suggest that this framework is realized in full general relativity. In particular, the notions of 4-momentum and angular momentum used in those analyses bear out the balance laws both for linear and angular momentum associated with $\inotP$ \cite{kkadl,kak}. As for analytical results, as we mentioned in section \ref{s2.4}, our boundary condition on the ``magnetic part" of the Weyl tensor is automatically satisfied in stationary \emph{or} axisymmetric space-times. Using the multipolar expansion that is available near infinity in stationary space-times, it should be possible to show that the continuity condition is also automatically satisfied. In dynamical space-times, on the other hand, there are relatively few results on global existence in full general relativity. Rigorous analysis to date refers to electrovac solutions \cite{dcsk,bieri,zipser,pced,kroon2} or solutions with null fluids as sources \cite{bieri2} on $\mathbb{R}^4$.  Precise fall-offs deduced from these analyses are sometimes assumed to hold also for compact binaries. However, strictly speaking, the assumptions underlying rigorous results to date preclude compact binary systems;  for these systems the issue of existence of asymptotically flat solutions satisfying remains open in full general relativity. 

(4) Nonetheless, these global results are clearly valuable in their own right, and they also provide intuition on the viability of proposals for boundary conditions that systems of more direct physical interest should satisfy, particularly at $\scrip$. For example, there is considerable literature on the viability of the $C^4$ differentiability assumption that leads to the Newman Penrose peeling behavior at $\scrip$ (see, e.g., \cite{dcsk,bieri,pced,hf-peel}). While there is a large class of initial data whose evolution leads to space-times in which $g_{ab}$ is $C^4$ \cite{pced}, there is larger class whose evolution yields lower differentiability at $\scrip$ (see, in particular \cite{dcsk}). However, even with this lower differentiability, ${}^\star{\Ko}_{ab}$ is well-defined at $\scrip$, i.e. the Newman-Penrose scalars $\Psi_4^\circ,\, \Psi_3^\circ$ and ${\rm Im}\Psi_2^\circ$ are continuous on $\scrip$. Therefore, our notion of \emph{AM} space-times can be weakened to allow this lower differentiability; an examination of our \Poincare reduction procedure shows that it will continue to go through. 
\medskip

(5) We will conclude with a clarification. There are two closely related notions of AEFANSI space-times in the literature each with its Spi framework. The essential differences are as follows. The first \cite{aarh} starts with a weakly asymptotically simple space-time and then introduces $\inot$ as the `vertex' of $\scri$. Motivated by the fact that, in these space-times, points of $\scri$ represent end points of null geodesics in the physical space-time, one considers curves in the completed space-time that are $C^{>1}$ at $\inot$ and space-like \emph{geodesics of the physical metric} $\hat{g}_{ab}$, with tangent vectors $\eta^a$ that are unit w.r.t. $g_{ab}$ at $\inot$. This $\eta^a$ captures the first order behavior of the curve at $\inot$. A careful examination shows that, while the component of the acceleration of this curve w.r.t. $\hat{g}_{ab}$ that is orthogonal to $\eta^a$ is fixed, its component along the curve is free and labels the `second order' behavior of these curves. One then considers a 4-dimensional fiber bundle $\B$ over the unit hyperboloid $\H$ spanned by the $\eta^a$ in $T_{\inot}$, where the 1-dimensional fibers are labeled by the tangential acceleration of the curves. The Spi group $\Spi$ is the group of automorphisms of $\B$. Now, given any one conformal completion, one can compute the tangential acceleration and introduce `horizontal' cross-sections of the bundle. Since each is naturally isomorphic to $\H$, one can lift the action of the Lorentz group $\L$ on $\H$ to $\B$. However, under allowed conformal transformations, $g_{ab} \to g^\prime_{ab} = \omega^2 g_{ab}$ where $\omega= 1 + \Omega^{\f{1}{2}}{\alpha}$, these cross-sections are shifted by an amount $\bfalpha (\eta) = \lim \alpha$. These shifts correspond to supertranslations, and we acquire as many Lorentz subgroups of the symmetry group as there are supertranslations. Hence, $\Spi$ has the structure of the semi-direct product $\S\, \ltimes \L$. 

The second formulation of the Spi framework \cite{aa-ein} moves away from the emphasis on geodesics both in the null and spatial regimes.  One first introduces $\inot$ --without any reference to $\scri$-- as the point in the completion $M$ that is space-like related to all points in the physical space-time $\hat{M}$. $\scri$ is then identified as the null cone of $\inot$ using the causality and time orientability conditions of \emph{Definition 3}. The first order structure consists of $\inot$ and the metric $g_{ab}\mid_{\inot}$. The second order structure is now encoded in the `ripples' --equivalence classes of asymptotic connections $\nabla$ where two are equivalent if their action agrees at $\inot$ (on all $C^{>0}$ tensor fields). As we discussed in section \ref{s3.1} this formulation is better suited for `gluing' together asymptotic flatness in null and spatial directions because it avoids the introduction of the 4-dimensional bundle $\B$. In this approach, the structures at forefront are the point $\inot$, the metric there, and the equivalence classes of asymptotic connections. Structures at forefront in the analysis of null infinity are the manifold $\scrip$, the intrinsic metric there and equivalence classes of connections. This similarity makes the framework well tailored for the unified treatment. That is why we used it in the main text.

\section*{Acknowledgments}

We thank Kartik Prabhu for discussions, Amitabha Veermani for his comments on the manuscript, and Gabriele Veneziano for motivating us to write up these results. AA also thanks Stephen  Hawking for a discussion in which he suggested that the asymptotic symmetry group for isolated systems in general relativity should be the \Poincare group. This work was supported in part by the Eberly and Atherton research funds of Penn State and the Distinguished Visiting Research Chair program of the Perimeter Institute. N.K. is supported by the Natural Sciences and Engineering Council of Canada.

\end{document}